\documentclass[12pt,preprint]{aastex}

\slugcomment{Accepted in The Astrophysical Journal}

\shorttitle{Low Surface Brightness Galaxies in the SDSS}
\shortauthors{Galaz et al.}

\begin{document}


\title{Low Surface Brightness Galaxies in the SDSS: the link between
  environment, star-forming properties and AGN} %%NELSON: como vos
                                %%decias, habria que poner AGN en el
                                %%titulo, que te parece? 


\author{Gaspar Galaz, Rodrigo Herrera-Camus\altaffilmark{1}}
\affil{Departamento de Astronomia y Astrofisica, Pontificia
  Universidad Catolica de Chile}
\email{ggalaz@astro.puc.cl, rhc@astro.umd.edu}
\author{Diego Garcia-Lambas}
\affil{Consejo Nacional de Investigaciones Cientificas y Tecnicas,
  Argentina, IATE, CONICET, OAC, Universidad Nacional de Cordoba, Argentina}
\email{dgl@mail.oac.uncor.edu}
\and
\author{Nelson Padilla}
\affil{Departamento de Astronomia y Astrofisica, Pontificia
  Universidad Catolica de Chile}
\email{npadilla@astro.puc.cl}


\altaffiltext{1}{Present address: Department of Astronomy, University
  of Maryland at College Park, USA.}
%\altaffiltext{2}{Society of Fellows, Harvard University.}
%\altaffiltext{3}{present address: Center for Astrophysics,
%    60 Garden Street, Cambridge, MA 02138}
%\altaffiltext{4}{Visiting Programmer, Space Telescope Science Institute}
%\altaffiltext{5}{Patron, Alonso's Bar and Grill}


\begin{abstract}
Using the Sloan Digital Sky Survey (SDSS) data release 4 (DR 4), we
investigate the spatial distribution of low and high surface
brightness galaxies (LSBGs and HSBGs, respectively). In particular, we
focus our attention on the influence of interactions between galaxies 
on the star formation strength in the redshift range $0.01 < z <
0.1$. With cylinder counts and projected distance to the first and fifth-nearest
neighbor as environment tracers, we find that LSBGs tend to have a
lack of companions compared to HSBGs at small scales ($<2$
Mpc). Regarding the interactions, we have evidence that the fraction of LSBGs
with strong star formation activity increases when the distance
between pairs of galaxies ($r_{p}$) is smaller than about four times the
Petrosian radius ($r_{90}$) of one of the components. Our results suggest
that, rather than being a condition for their formation, the
isolation of LSBGs is more connected with their survival and
evolution.   
The effect of the interaction on the star formation strength,
measured by the average value of the birthrate parameter $b$, seems to
be stronger for HSBGs than for LSBGs. The analysis of our population of
LSBGs and HSBGs hosting an AGN show that, regardless of the mass range,
the fraction of LSBGs having an AGN is lower than the corresponding
fraction of HSBGs with an AGN. Also, we observe that the fraction of
HSBGs and LSBGs having an AGN increases with the bulge
luminosity. These results, and those concerning the 
star-forming properties of LSBGs as a function of the environment,
fit with the scenario proposed by some authors where, below a
given threshold of surface mass density, low surface brightness disks
are unable to propagate 
instabilities, preventing the formation and evolution of massive black
holes in the centers of LSBGs.
\end{abstract}





\section{Introduction: new questions}

Low surface brightness galaxies (LSBGs hereafter) represent an important
population among extragalactic objects. In particular, spiral LSBGs are
characterized by a disk surface brightness at least one order of magnitude
lower than the canonical value of 21.65 mag arcsec$^{-2}$ proposed by
\citet{freeman1970}. The central surface brightness of the disk in the
$B$-band, $\mu_{0}(B)$, is the photometric parameter typically used to
distinguish between the high and the low surface brightness regime of
galaxies. The most common threshold values found in the literature
are between 22 and 23 mag arcsec$^{-2}$ \citep[among
others]{impey2001}. Although LSBGs share many of the properties also
found in high surface brightness galaxies (HSBGs), they also present a
significant list of challenging observational features. Just to
mention the most intriguing ones (some of them will be explained more
extensively in the next paragraph), they have a very low stellar density
(which actually produces 
the low surface brightness), but exhibit astonishing flat rotation
curves, reaching  large radii from the center of the galaxy
\citep{deblok2005, swaters2010}. This implies that LSBGs are 
one of the most dark matter dominated systems in the Universe, given
their high M/L ratio \citep{sprayberry1995a}. Another striking feature in LSBGs is 
the richness of their stellar populations, which span the
whole range of the HR-diagram \citep{zackrisson2005,
  zhong2008}, challenging 
the extraordinary deficit in molecular gas, as detected so
far \citep{oneil2000, matthews2001, galaz2008}. Finally, the low
star formation rate (SFR) in combination with 
their rather isolated location in the cosmic web
\citep{rosenbaum2009}, as reported by 
several authors (see below), give clues for the understanding
of their formation and evolution. 
%All of this turn LSBGs and in
%particular the efforts to disentangle their unclear nature, into an
%interesting and boiling field of research. 

Among the mentioned properties, we highlight three 
of them related with the purpose of this work. First, it seems that
LSBGs evolve following a similar track of high surface brightness
galaxies (HSBGs), but with a significantly slower rate of star forming
processes \citep{hoek2000}. Second, systematic evidence shows that
LSBGs are strongly  dominated by dark matter \citep{deblok1996}. Then,
far from being fragile, simulations suggest that disks of LSBGs would be very
stable against the propagation of gravitational instabilities which
are typically generated by interactions with, for example, a close neighbor
\citep{mihos1997, mayer2004}. Finally, LSBGs would be more isolated than HSBGs at
small scales (less than 2 Mpc), according to \citet{bothun1993}, and
between 2 Mpc and 5 Mpc \citep{rosenbaum2009}. These last results
motivate the following question: is the isolation of LSBGs a requisite for their
formation or for their subsequent evolution? In other words, if the
reaction of LSBGs to close interactions results in the enhancement of
their star formation activity, as has been previously observed in HSBGs
\citep{lambas2003, nikolic2004}, then the lack of tidal  
encounters in a Hubble time, because of their isolation, could explain
the less evolved nature of LSBGs respect to HSBGs
\citep{bothun1987}. On the other hand, if LSBGs are strongly influenced by the 
interactions, then the isolation would indicate a fast transition from
an LSB regime to the HSB regime in high density
environments. 
%Furthermore, if LSB disks are stable against
%interactions, then their isolation would be related 
%with their formation rather than their evolution.  

In an effort to better understand the relationship between the
spatial distribution of LSBGs and their star formation properties, we
focus our attention in a carefully selected sample of this kind of
galaxies extracted from the Sloan Digital Sky Survey (SDSS) data release 4 (DR 4)
\citep{abazajian2004}, as well as in a similarly selected control sample of
high surface brightness galaxies from the same catalog. Along with the
analysis of the degree of isolation of LSBGs, we will focus our attention
on the possible relationship between the local density and their
star-formation properties, including the behavior of LSBGs in
pairs. 

The paper is organized as follows. In \S\ref{sample} we describe the
galaxy selection from the SDSS DR 4 and the corrections applied to
such a selection, showing also some basic relationships;
in \S\ref{density} we define the density estimators to be used in the
analysis, as well as the tools to define galaxy pairs. In
\S\ref{results} we analyse the most important results regarding the
environment and the star-formation properties of LSBGs, including an
analysis of the LSB population exhibiting an AGN. We discuss our
results within the scope of the recent work by other 
authors in
\S\ref{discussion}. Finally, we summarize our conclusions in
\S\ref{conclusion}.      

\section{The sample}
\label{sample}

\subsection{Selecting low surface brightness galaxies from
  the SDSS}
\label{selecting}

The galaxies studied in this work were extracted from the Main Galaxy
Sample \citep{strauss2002} of the SDSS data release 4 (DR 4). Following a
similar procedure to that presented 
by \citet{zhong2008}, we select preferentially late
type galaxies 
($fracDev_{r} \leq 0.9$)\footnote{The {\em fracDev}
  parameter in the SDSS refers basically to the fraction of the light
  coming from a fitted de Vaucouleurs light profile.} with spectroscopically
computed redshifts, nearly face-on to avoid serious extinction
correction ($b/a > 0.4$), not too nearby to avoid problems with peculiar
velocities ($z \geq 0.01$) and, to avoid serious incompleteness
effects, not too distant ($z \leq 0.1$, see also Rosenbaum et
al. 2009). 
Note that \citet{zhong2008}, using also the SDSS, selected
galaxies with $fracDev_{r} \leq 0.25$. In spite that this last figure
seems to be quite different than our selection criteria, we realize
that most of the LSBGs included in 
our sample (86\%) do have  $fracDev_{r} \leq 0.25$. It is worth noting
that recently, \citet{rosenbaum2009} did an excellent work selecting
LSBGs from the same catalog (DR 4), and concluded that the SDSS is
biased to select two different kinds
 of galaxies in two different
redshift intervals, $0.010 < z < 0.055$ and $0.055 < z < 0.100$, which
span our redshift selection. In the first redshift range, most of the LSBGs
are dwarf-type galaxies, LMC-like. In the second redshift range,
the majority of galaxies are massive, well defined spiral systems
(where in fact one can very well define a disk and bulge). With our
luminosity profile selection ($fracDev_{r} \leq 0.90$) we are only
loosing some extreme LSBGs with a huge bulge component and very faint
disks (i.e. galaxies where the de Vaucouleurs component contributes
more than 90\% of the total light). On the other hand, irregular
galaxies enter in our catalog if an exponential light profile is
possible to be fitted, which is the case for about 95\% of the cases
in the spectroscopic catalog. Therefore, comparisons between this work and the
one by \citet{rosenbaum2009} can in principle be made. In fact,
further in the text it is possible to conclude that our results
regarding the large-scale distribution and the distribution of LSBGs in
terms of the local density, agree well with Rosenbaum et al..

Since the spectroscopic catalog of the SDSS is a magnitude-limited
survey ($r \le 17.77$ mag), the observed populations of galaxies are not the
same as the redshift increases. In fact, when going farther in
redshift, a magnitude-limited selection introduces a strong bias
selecting more luminous (and massive) galaxies. This
prevents us to compare in a statistical way nearby galaxies with galaxies
situated at higher redshifts. Therefore, one has to define a redshift
range (i.e. a volume and a lower luminosity limit) where the absolute
magnitude distributions are the same for all the galaxy populations.
A volume-limited catalog allows us to compare two populations of
LSBGs and HSBGs having roughly the 
same absolute magnitude range and therefore with roughly the same mass
cuts, as we show in furthers sections. Another way to correct for the
bias from a magnitude-limited catalog is to use the $V/V_{max}$
approach, which will also be used (see the end of
\S\ref{volume_correction}).

\subsection{Magnitude conversion and surface brightness}
\label{magnitude}

Since the criterion for LSB classification is done in the Johnson $B$
band, and the SDSS filters does not include it,
we are forced to transform the SDSS $g$ and $r$ magnitudes into $B$-Johnson
magnitudes. 
Using information derived from the fit of a pure exponential
profile included in the SDSS data, we calculate the surface brightness in the
$g$- and $r$-band. Then, from the conversions of \citet{smith2002}, we
get the surface brightness in the $B$-band, which is computed as 

\begin{equation}
\mu_0(m) = m + 2.5 log(2\pi a^2) + 2.5 log (b/a) - 10 log(1+z),
\label{mu_def}
\end{equation}

\noindent where $m$ is the apparent magnitude in the $B$ band, the second
factor is the area where the light is measured 
(enclosing 90\% of the light), the third factor is the inclination
correction (where $a$ is the semi-major axis, $b$ is the minor axis)
assuming a disk of uniform radiance, and the fourth factor 
is the surface brightness correction for the cosmological dimming (the
well-known $(1 + z)^{-4}$ factor). The $a$ and $b$ parameters are defined
from the photometric exponential fit to the galaxy 
%(expAB in the PhotoObj table) 

Using a surface brightness cut value of $\mu_{0}(B) = 22.5$ mag
arcsec$^{-2}$, we 
select a sample of 9,421 LSBGs from the complete spectroscopic catalog
of 567,486 galaxies, i.e. 1.66\% of the galaxies of the total sample
are LSBGs. These galaxies satisfy the
surface brightness cut and also the redshift, inclination and
morphology constraints specified in \S\ref{selecting}.  
Applying the same redshift cuts and imposing the  
same spectroscopic flags as for the LSB sample, we select 30,000 HSB
galaxies.
%The color of both populations can be observed in Figure
%\ref{lsbs_hsbs}. 
In addition to the Sloan spectroscopic information,
we have added information from \citet{brinchmann2004}, which includes
the star formation rate, the estimated stellar mass for each
galaxy ($M_*$) and the 4000 \AA ~ break index (D$_n$4000). Also, with
the aim to flag galaxies hosting an active galactic nucleus (AGN), we
cross-correlate our catalog with that of \citet{kauffmann2003}. These
galaxies have to be excluded from our analysis when studying the light
and colors supposedly coming from canonical star forming
processes. Therefore, if one excluded galaxies with AGN, then the
number of LSBGs is reduced to 8,926 and the number of HSBGs to
24,324\footnote{The HSB number of galaxies is only a
sub-sample of the total sample build as a control sample.}. Specific
studies of AGN are presented in a forthcoming  
paper (however, see \S\ref{agn}). 

\subsection{Volume corrections}
\label{volume_correction}

\begin{figure}
\plotone{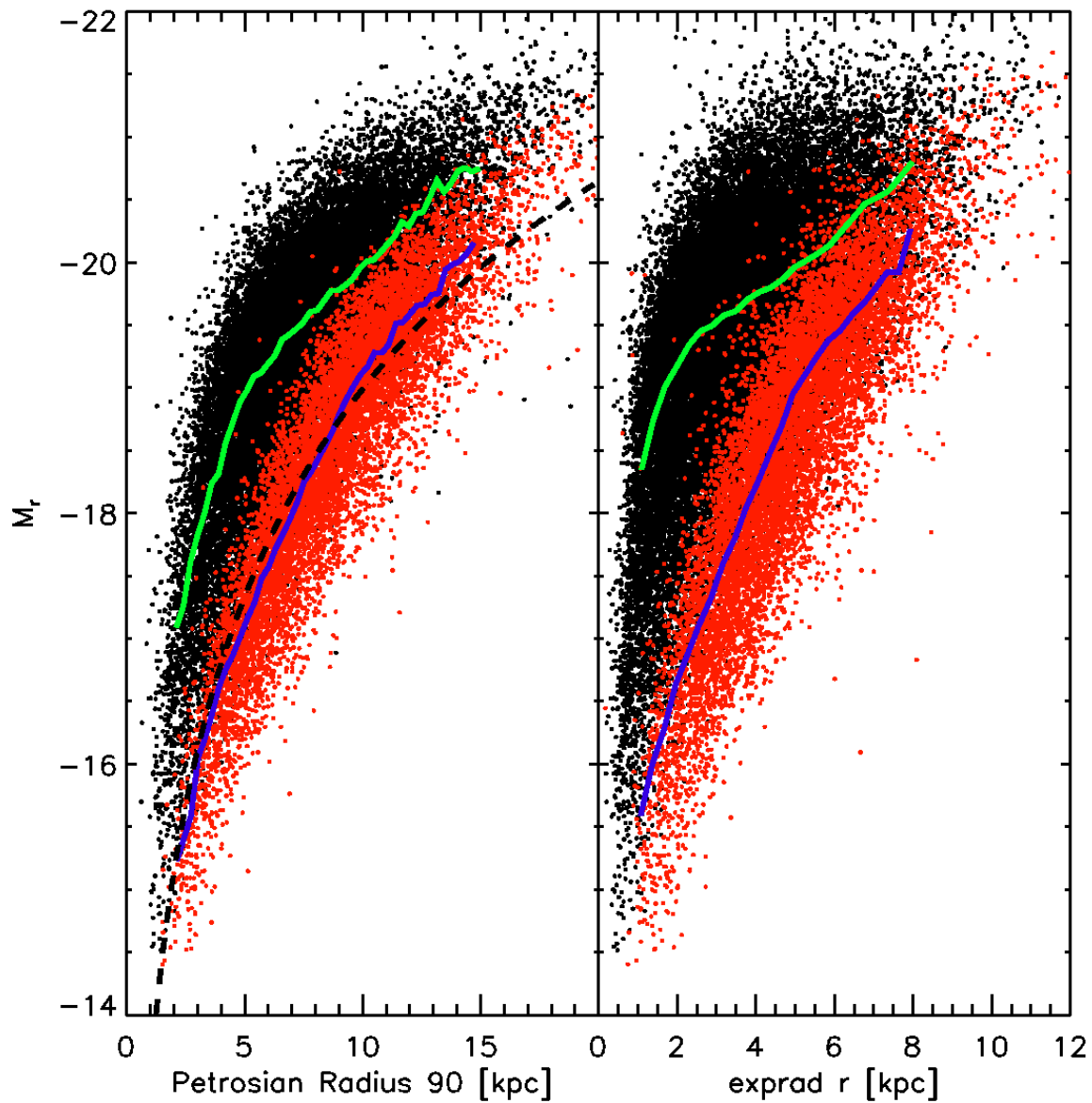}
\caption{In black HSBGs and in red LSBGs. Left panel:
  Absolute magnitude in the $r$ band vs. the Petrosian radius 
  including 
 90\% of the light. Right panel: the same absolute
  magnitude but as a function of the exponential scale length measured
  in kpc. Black dashed line indicates the region where 90\% of the
  galaxies are LSBGs; Green and blue lines correspond to the median for
  HSBGs and LSBGs, respectively. \label{petro_abs}}
\end{figure}

In order to correct the bias
introduced from the differences in both the redshift and absolute magnitude
distributions for the HSBGs and LSBGs, we
weight the fractions and averages with the inverse of the maximum
volume out to which each galaxy can be detected in the SDSS (i.e. we use the
$V/V_{max}$ method, \citealt*{schmidt1968}). 
%The total number of LSBGs
%is 9,241 and the total 
%number of HSBGs is 30,000 (N.B.: The HSB number of galaxies is only a
%sub-sample of the total sample build as a control sample). 
%For the spatial distribution analysis, we use
%volume-limited samples ($M_{r} \leq -19.8$). 
Figure \ref{petro_abs} shows the relationship between the absolute
magnitude, the Petrosian radius $r_{90}$\footnote{The equivalent
  circular radius enclosing 90\% of the galaxy light, in the
  $r$-band.} and the exponential 
scale-length, for HSBGs (in black) as well as for LSBGs (in red). This
Figure clearly presents the combined bias resulting from two selection
functions, namely, the strong dependence of the absolute magnitude
vs. redshift in the SDSS spectroscopic catalog (a magnitude-limited
catalog), and the trend of the absolute magnitude on the galaxy
size. By weighting each galaxy for its accessible volume using the
factor $V/V_{max}$, we correct the effective fractions of galaxies,
specially the few ones detected at faint absolute magnitudes (the bottom edge in
Figure \ref{petro_abs}). Figure \ref{petro_abs} clearly
shows that for a given absolute magnitude, LSBGs are larger than
HSBGs. Note that our sample of galaxies also includes some irregular
galaxies, but not the usually faint ones. These have small values of
fracDev ($< 0.2$) and faint absolute magnitudes (usually $M_r > -16.0$,
see Figure \ref{petro_abs}). Given that our spectroscopic catalogue is a 
magnitude limited catalog with with $m_r \le 17.77$ mag, these galaxies
are visible only up to $z \sim 0.01$, which is exactly the
redshift we start to consider for this study. Then we do not
detect these faint irregulars. To have an idea
about how many irregulars are we loosing up to such a redshift we
first selected the total number of LSBGs and HSBGs up to $z = 0.01$, which
gives 741 and 1099 respectively. From these, about 30\% are galaxies
of small Petrosian radius (smaller than 2 kpc). 
%as shown in Figure \ref{low_z}). 
This fraction is about the same for the low surface
brightness and the high surface brightness population, as shown by the
solid and dashed lines of the histogram. If irregulars
are represented by these galaxies, that means that we are loosing a
40\% of the total sample, which is a significant fraction. However,
only 30\% of them are fainter than $M_r < -16.0$. Also, it
is well known that there are also LSBGs 
with small radius and not necessarily irregulars, for example dwarf
galaxies that enter 
 into our 
sample \citep{delapparent2004}. So the fraction of irregulars is
probably smaller than 40\%.   
%These corresponds to 7\% and 1\% of
%the total number of LSBGs and HSBGs up to $z = 0.1$,
%respectively. %NEW ----->>>> 
To be sure, we have examined in detail the number of irregulars at $z
\le 0.01$ that are included in the sample of 741 LSBGs up to $z \sim
0.01$. We found, {\em by eye}, that 127 (17\%) of these are clearly
irregulars. A significant number of galaxies, about the same
fraction, are dwarf systems with sizes $\sim 15$ kpc or smaller.  
%%%%%%%%%%%

\begin{figure}
\plotone{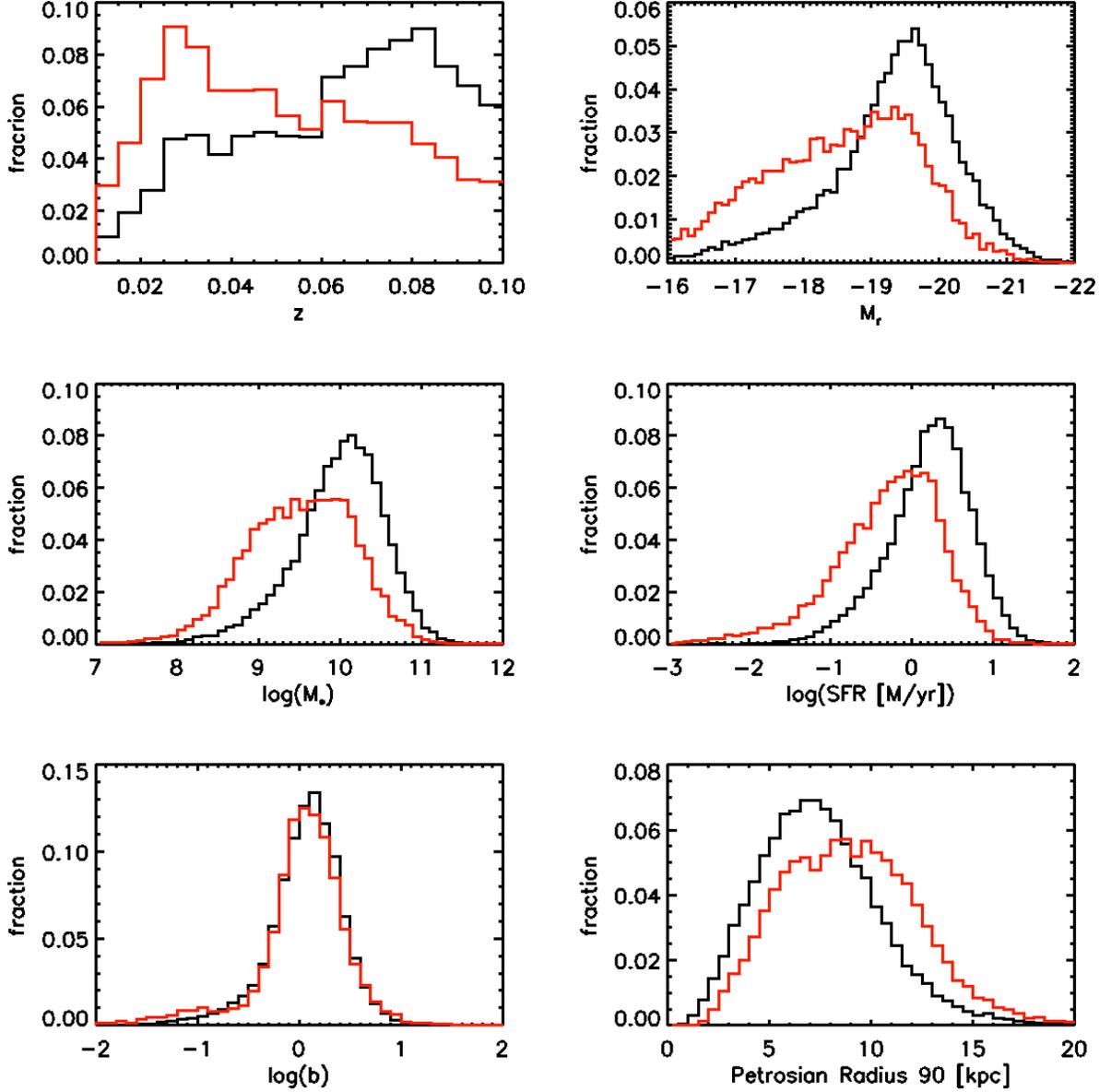}
\caption{$V/V_{max}$ weighted distributions of galaxy properties for
  our sample of LSBGs (red)
  and HSBGs (black). From up to bottom, from left to right: the redshift ($z$)
  distribution, the absolute magnitude in the $r$ band, the log of the
  stellar mass (M$_*$), the log of the star formation rate, in
  solar masses per year (M$_\odot$/year), the log of the birthrate
  parameter $b$, and the Petrosian radius with
 90\% of the
  galaxy light ($r_{90}$).  The observed distributions have been
  corrected from volume incompleteness by the factor $V/V_{max}$. See
  text for details. \label{gen_props}}  
\end{figure}

Figure \ref{gen_props} shows
the distribution of properties extracted from our
sample of LSBGs and HSBGs, considering the $V/V_{max}$ weight. Looking
at the two upper histograms, we observe 
indeed that LSBGs (in red) and HSBGs (in black) do not have the same
redshift and absolute magnitude distributions. This Figure is consistent
with Figure 1 in \citet{rosenbaum2009}. 
%The difference is
%originated from convolution of the 
%magnitude limited catalog (as it is the case of the spectroscopic
%catalog, limited to galaxies brighter than $r = 17.77$ mag), with the
%surface magnitude-absolute magnitude relationship (see Figure
%\ref{petro_abs}). 
The middle panels in Figure \ref{gen_props} show the
fractional histograms for the stellar mass and the star formation
rate, for HSBGs and LSBGs, weighted by the  $V/V_{max}$
factor. We see that LSBGs peak at slightly less 
massive objects compared to HSBGs, and also with a broader mass
distribution. A similar result is observed for the SFR distribution
(right panel), where the LSBGs peak at about 0.9 M$_\odot$/yr,
compared to HSBGs, which in average form more stars per year, 1.8
M$_\odot$/yr, as expected. The two bottom panels show the distributions
of the birthrate parameter $b$ and the Petrosian radius,
respectively. The birthrate parameter is defined as 

\begin{equation}
b = (1 - R)t_H\frac{SFR}{M_\star},
\end{equation}

\noindent where $R$ is the total stellar mass fraction which is ejected to the
interstellar medium (ISM) during the lifetime of the galaxy, $t_H$ is
the Hubble time, and the ratio $SFR/M_*$ is the star formation rate (SFR)
per unit of solar mass (also called specific SFR\footnote{The specific
  SFR is usually a better estimator than the SFR for the actual amount of mass
  which forms stars in a given galaxy, since it is the SFR per unit of
  galaxy mass.}). Usually $R$ is set as 0.5
\citep{brinchmann2004}. Interestingly, Figure \ref{gen_props} shows that LSBGs and HSBGs
have the same birthrate parameter distribution, when corrected by the
$V/V_{max}$ factor. Also, the right histogram shows that LSBGs and HSBGs
have clearly different Petrosian radius $r_{90}$ distributions, both
in range and peak values. 

\subsection{Volume-limited sample}
\label{volume-limited}

\begin{figure}[!ht]
\plotone{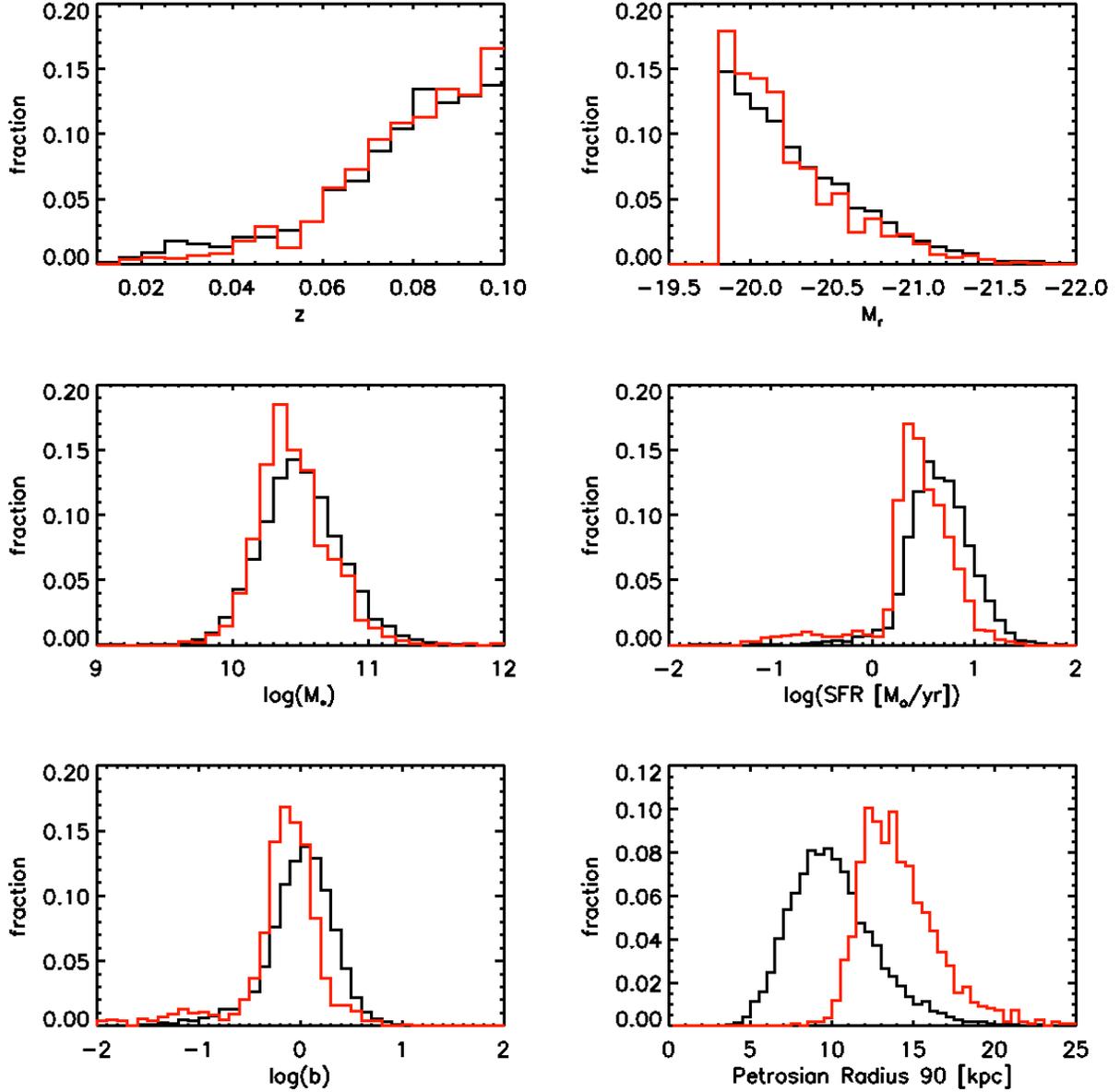}
\caption{In red, LSBGs; in black, HSBGs. Same as Figure
  \ref{gen_props} but now considering only the 
  volume-limited catalog. We see that the redshift and absolute magnitude
  distributions are heavily modified. See text for
  details. \label{gen_props_corrected}}   

\end{figure}

All the distributions shown in Figure \ref{gen_props} are slightly
modified if one directly uses a volume-limited sample of galaxies,
instead of the
$V/V_{max}$ weighted distributions. However, even though the absolute
magnitude-redshift bias is eliminated, the number of galaxies is
small. Another drawback of having a volume-limited sample is that the
catalog now includes preferentially bright galaxies, in our case
galaxies brighter than $M_r = -19.8$ mag. When this cut in absolute
magnitude is used, the total number of LSBGs is
reduced to 1,110, and the corresponding number of HSBGs to 7,526.   
%corrects for the volume access each galaxy really has, in order to
%have two samples comparable statistically. 
Figure \ref{gen_props_corrected} shows the same histograms as Figure
\ref{gen_props} but for the volume-limited sample.
%but corrected by
%the $V/V_{max}$ factor. 
We see that most of them are consequently
modified, especially the redshift and absolute magnitude
distributions, as expected. The middle panel of Figure
\ref{gen_props_corrected} shows that now the two populations do have
more similar stellar mass 
distributions (easily understood from the now similar absolute
magnitude distribution), although LSBGs still have lower masses. 

Also, now the birthrate distributions for LSBGs and HSBGs
are no longer the same. Indeed as expected, LSBGs form in average, less
solar masses per year, during their lifetime. That could be
interpreted as the fact that although LSBGs and HSBGs could have the same
stellar formation history, LSBGs form stars in a longer periods of time. 

Finally, the right bottom panel indicate that the Petrosian radius
distributions are not only still different, as expected from the definition
itself of low surface brightness, but are markedly different, both in
peak values and in range. 

Table \ref{stat_props} shows the averaged quantities studied
in Figure \ref{gen_props} and \ref{gen_props_corrected} for LSBGs as
well as for HSBGs. In 
average, LSBGs with the same average total absolute magnitudes
of the HSB counterparts, have
the same stellar mass, but form 8 times less solar masses per year, and are
1.5 times larger than HSBGs. Note also that the average $b$ parameter
is higher in HSBGs than in LSBGs, for the two ways of weighting
galaxies. In particular, for the volume limited sample, LSBGs form
almost 10 times less stars per unit mass than HSBGs. 

\section{Density estimators and pair recognition}
\label{density}

In this section we briefly define the density estimators that will be used to
analyze the degree of isolation of our sample of LSBGs. We also define
the parameters and criteria to define our sample of pairs.  

We use two methods to estimate the spatial density of galaxies: (1)
the object counting in a fixed aperture over the sky, and (2) the distance to the
n$^{th}$-closest neighbor for a given galaxy. In both cases, we only use the
galaxies included in the volume-limited sample ($M_{r} \leq -19.8$). 

In the first method, also used by \citet{bothun1993} for
340 LSBGs, and recently by \citet{rosenbaum2009}, we count the number of
galaxies in cylinders of fixed height ($\pm500$ km/s) 
and radii up to 5 Mpc in steps of 0.25 Mpc, to calculate the
surface density of galaxies inside each cylinder. For a given
galaxy the number of neighbors included in each cylinder allows to
define a surface density of galaxies as a function of radius, which in turn
can be used to define its local density.

In the second method, we compute both the projected distance to the nearest
neighbor $r_p$ (i.e. $N=1$) within a velocity shell of $\pm500$ km/s, and
the projected distance to the fifth nearest neighbor $d_5$ ($N=5$), brighter
than the K-corrected absolute magnitude M$_{r} = -19.8$,  and within a
velocity range of $\pm 500$ km/s. This last estimator allows us to
characterize the local environment around the target galaxy by means
of the well-known surface density parameter $\Sigma_5$ \citep{rosenbaum2004,
  balogh2004, alonso2006, padilla2010} 
 defined by 

\begin{equation}
\Sigma_5 = \frac{5}{\pi d_5^2},
\end{equation}

\noindent where $d_5$ is the projected distance to the 5$^{th}$-closest and
brighter than $M_r = -19.8$ mag neighbor. 

%NELSON: Esto lo saco porque hace dudar de estas medidas de ambiente que 
%        han sido mas que probadas
%We believe that in spite of the fact that these are simple tools,
%they can provide us with clean and enough information to determine the
%degree of isolation of LSBGs.  

\section{Results}
\label{results}
\subsection{How isolated are LSBGs? Environment around LSBGs}
\label{isolated}

\begin{figure}[!ht]
\plotone{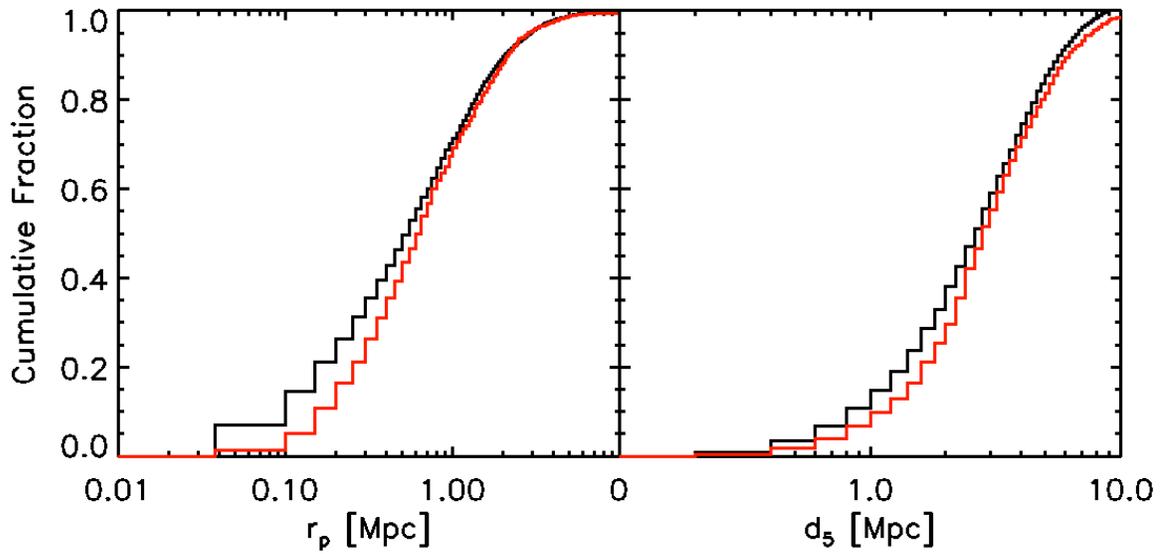}
\caption{Cumulative distribution of the first (left panel) and fifth
  (right panel) nearest neighbor, for LSBGs (red) and HSBGs
  (black). Note that LSBGs always have a smaller number of neighbors
  than HSBGs, at all scales, but more significantly for scales smaller than
  $\sim 4$ Mpc for $d_5$. See text for a detailed discussion. \label{d1_d5}}  
\end{figure}

In this section we analyze the spatial distribution of the
volume-limited sample of LSBGs and HSBGs, using the two estimators
defined in \S\ref{density}. 
Figure \ref{d1_d5} shows the cumulative distribution of $r_p$ (left
panel) and $d_5$ (right panel) for both LSBGs (red lines) and HSBGs
(black lines). It is apparent from this Figure that for both neighbor
definitions, LSBGs are more isolated than HSBGs, specially at scales
smaller than 5 Mpc. Indeed, at a fixed cumulative fraction, LSBGs tend to have
the fifth nearest neighbor {\em farther} away than HSBGs, in the region
%$0.5 \leq r \leq 3.0$ Mpc. 
%NELSON: para toda escala!  el rp si tiene una escala maxima pero si
%ves las lineas negra y roja a 10Mpc (derecha) estan muy separadas.
%Cambiamos a todo el rango para d5 y hasta 2Mpc para rp como sigue?: 
$0.5 \leq r\leq 2$Mpc for $r_p$ and almost the full range of scales
shown for $d_5$ ($0.5\leq r\leq 10$Mpc). 
%FIN
A Kolmogorov-Smirnov (K-S) test rejects,
with high level of confidence ($99\%$), the hypothesis that the
distance distribution is the same for both samples of galaxies. 
This behavior is still more apparent for scales
smaller than 2 Mpc. Note that we do not observe a significant degree of
isolation in LSBGs at intermediate scales as described by
\citet{rosenbaum2004, rosenbaum2009}, probably because we are using a
volume-limited sample.
%for volume limited samples. 

\begin{figure}
\plotone{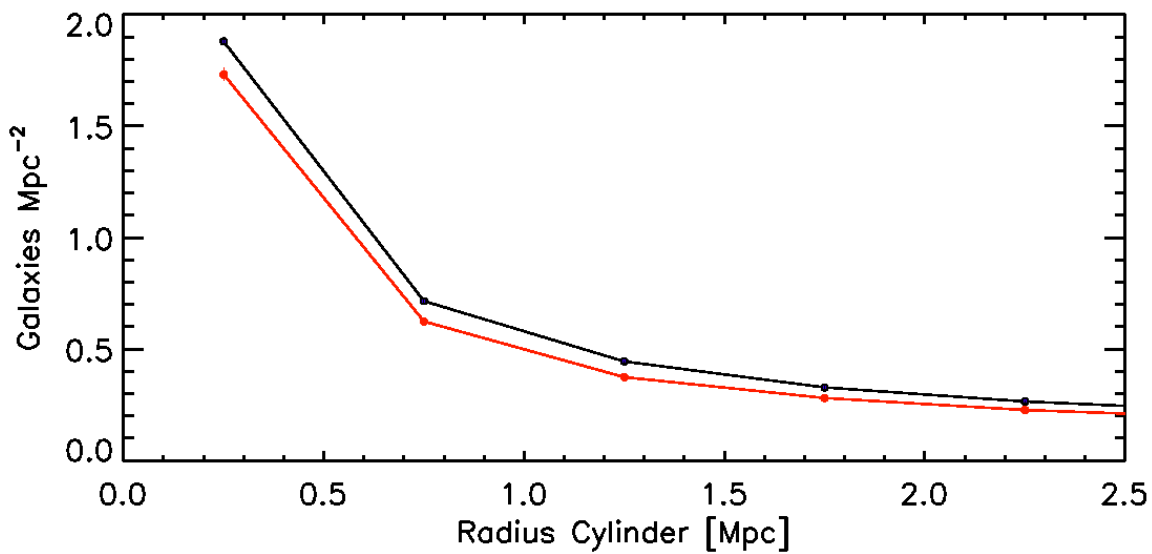}
\caption{Surface density of neighbors around LSBGs (red) and HSBGs (black) as a
  function of the radius of the cylinder. Note that error bars are of
  similar size as
 the symbol size. This estimator also indicate that
  LSBGs are more isolated than HSBGs. \label{surface_density}}
\end{figure}

When using the cylinder estimator one gets the same results as
with the nearest first and fifth neighbor. 
%Figure \ref{cylinder_counts} shows the number of neighbors as a function of
%increasing cylinder radii. Here we also observe that LSBGs (red
%points) lie below HSBGs (black dots), i.e. for all the scales shown in
%the Figure (up to 2.5 Mpc).
A better estimator of the degree of isolation of galaxies, other than the
pure cylinder counts, is the surface density of
galaxies around a given target galaxy ($\Sigma_5$). This estimator is
shown in Figure \ref{surface_density}, where it is apparent that LSBGs (red)
have a lower number of companions than HSBGs (black), specially
% NELSON: agregue los colores de LSBGs y HSBGs 
at scales smaller than 1.5 Mpc. The percentage of galaxies without
neighbors at $r < 0.5$ Mpc is $76\pm2\%$ for LSBGs and $70\pm1\%$ for
HSBGs. Note that high density environments are not exclusively
inhabited by HSBGs. Nearly $7\%$ of LSBGs have 8 or more neighbors at $r
< 2.0$ Mpc, compared to $\sim10\%$ for the HSB case. 

Note that all the above indicators show that on average LSBGs are more
isolated than HSBGs. However, it is worth noting that although the last
affirmation is correct, it is also true that some LSBGs are also located
in dense environments, in particular in groups and pairs, as will
be discussed in \S\ref{interactingLSBGs}. 

With the aim to characterize the extremes of the local density as
populated by LSBGs, we follow the same convention as \citet{bothun1993}, who
defined {\em isolated} galaxies as those having no neighbors up to
a distance of 2 Mpc. On the other hand, galaxies having 8 or more
neighbors up to the same radius of 2 Mpc, are defined as
{\em populars}. Using this criterion, 14\%$\pm$2\% and 18\%$\pm$2\% of
LSBGs in our sample are classified as isolated and  populars,
respectively. Interestingly, a similar fraction of isolated HSB galaxies
is found (13\%$\pm$1\%). However, the fraction of popular galaxies for
HSBGs is clearly larger than for LSBGs: 23\%$\pm$1\%. Therefore, this approach
indicates that LSBGs are not necessarily more isolated than HSBGs, but
less ``sociable'' at scales smaller than 2 Mpc. Our results are in
excellent agreement with those early results by \citet{bothun1993} for their
small sample of 340 LSBGs.

\subsection{Properties of star-forming LSBGs as a function of the
  environment: LSBGs in pairs }
\label{interactingLSBGs}

Among the many issues regarding the formation and evolution of
LSBGs, their (deficient) star forming properties as a function of the
environment is one of the most unexplored. The conclusion of
\S\ref{isolated} could hold clues about the elusive connection between the 
local density and the star-formation properties of LSBGs. From Table
\ref{props_sfr}, which includes 
averaged values for the absolute magnitude, color, the birthrate
parameter $b$ and D$_n$4000, we are unable to conclude that environment
variations can produce a significant difference on the stellar formation
activity between LSBGs and HSBGs, at least on 
 the scales included in the
Table. It is worth to note that we have excluded AGN galaxies from the
analysis, using the AGN catalog by \Citet{kauffmann2003} (see
\S\ref{agn}). If the environment has a different influence on the stellar
formation activity in LSBGs and HSBGs, such an influence should be seen at 
small scales, where interactions between galaxies
take place. 

\begin{figure}
\plottwo{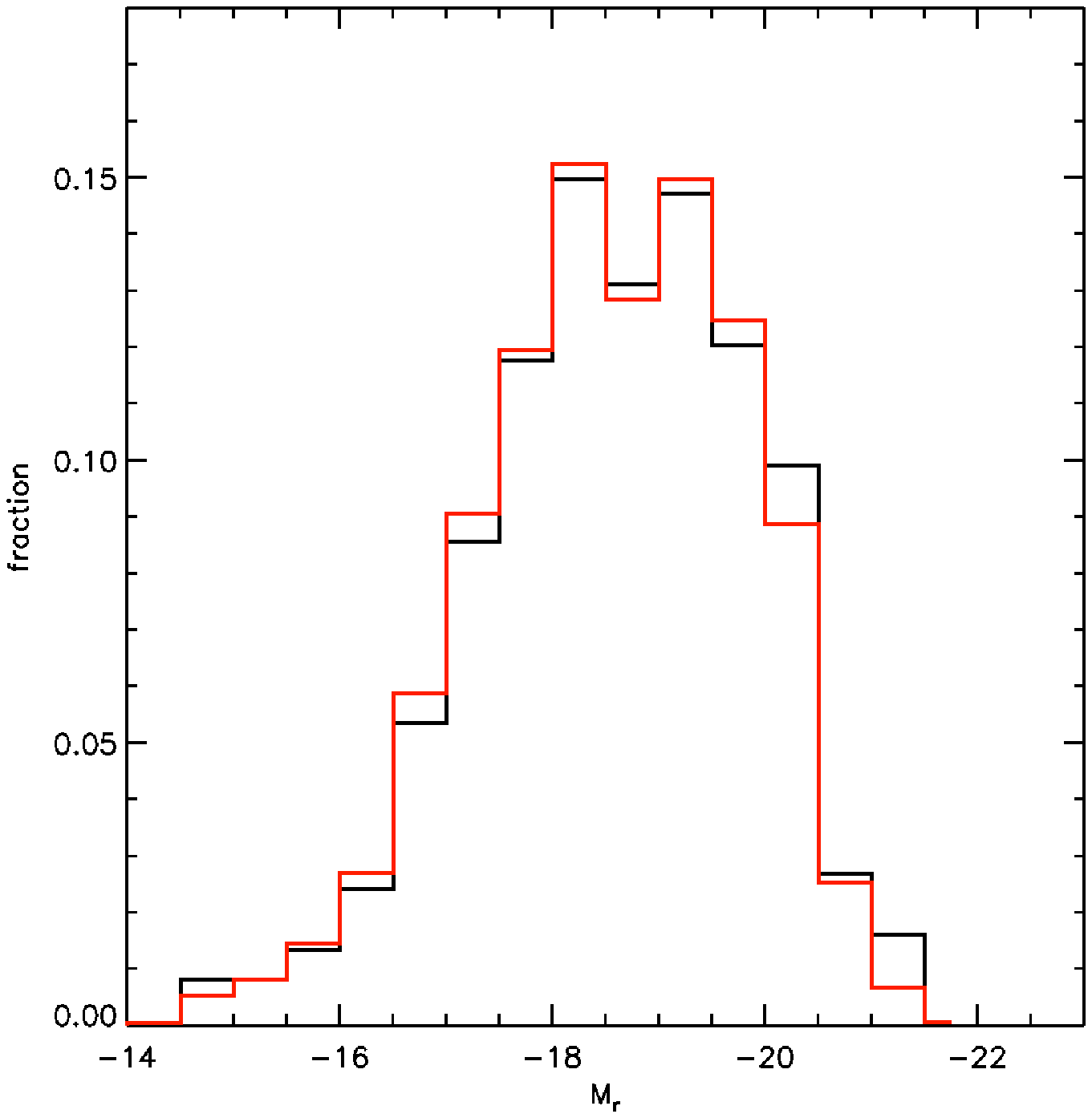}{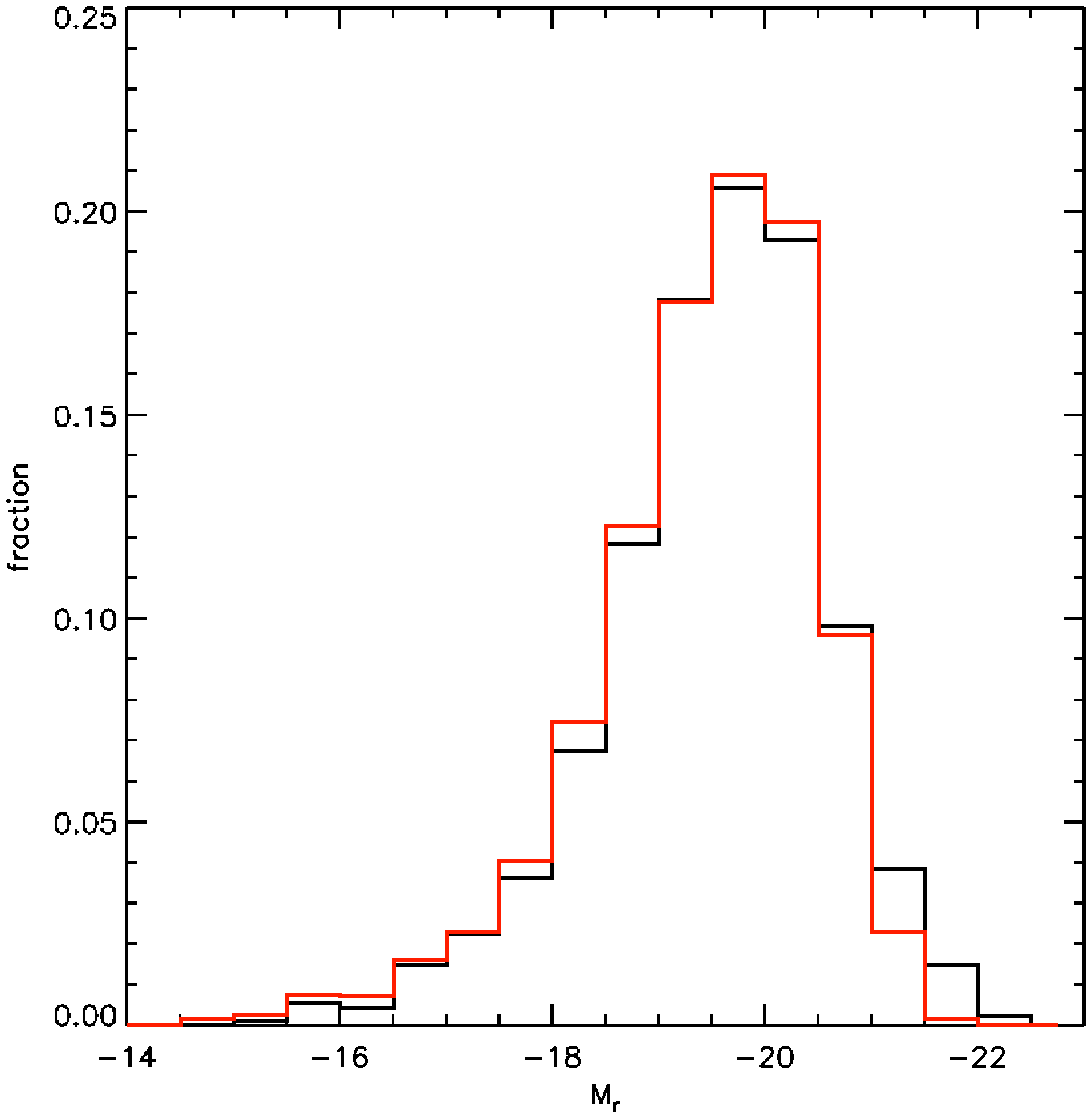}
\caption{Absolute magnitude distributions for interacting LSBGs (left
  panel, black line) and interacting HSBGs (right panel, black
  line). The red
lines correspond to the absolute magnitude
  distributions for the control non-interacting galaxies. See text for
  details about how these samples are built. \label{abs_mag_pairs}}
\end{figure}

To investigate the impact of interactions on 
the star forming properties on both LSBGs and HSBGs, we built a
sample of pairs from the main LSB and HSB catalogs. The pairs sample
is built using limits on the radial velocity differences ($\Delta V$)
between the central galaxy and its first 
neighbor, and the projected distance in Mpc ($r_p$) normalized
by the galaxy radius (here $r_{90}$, in kpc in the $r$ band). This is the same
approach as used in 
\citet{lambas2003}, and \citet{alonso2006}, allowing to
measure the distance to the corresponding companion in terms of the
influence radius of a galaxy. We note that the close neighbor of a given target LSB or HSB
galaxy is extracted from the whole spectroscopic SDSS database, not
necessarily included in our constrained catalogs of LSBGs and
HSBGs. This means that the 
neighbor can have any morphology, surface brightness, color or
magnitude (both apparent or absolute). The catalog of pairs
is composed by galaxies 
that satisfy, along with the neighbor galaxy, only two constraints:
$\Delta V \le 500$ km/s, and $r_p/r_{90} \le 10$. The number of pairs
selected for the sample without cuts in absolute magnitude turns out
to be 
268 and 868 for LSBGs and HSBGs, respectively. If we only consider
galaxies with a cut in absolute magnitude (i.e. galaxies with $M_r \le
-19.8$ mag) then the number of pairs where the target galaxy is an LSB
decreases to only 67 pairs. This number is too small to extract
statistical properties. Therefore, for this analysis we use the
complete sample of LSBGs and HSBGs, without any cut in absolute
magnitude, but correcting statistically the counts by the $V/V_{max}$
weight. We also built control samples 
from the galaxies not included in the pair selection criteria, as
counterparts for both the LSBGs and HSBGs pairs; the control galaxies
have the same absolute magnitude distribution as the those in
pairs (see Figure \ref{abs_mag_pairs}).  

\begin{figure}
\plotone{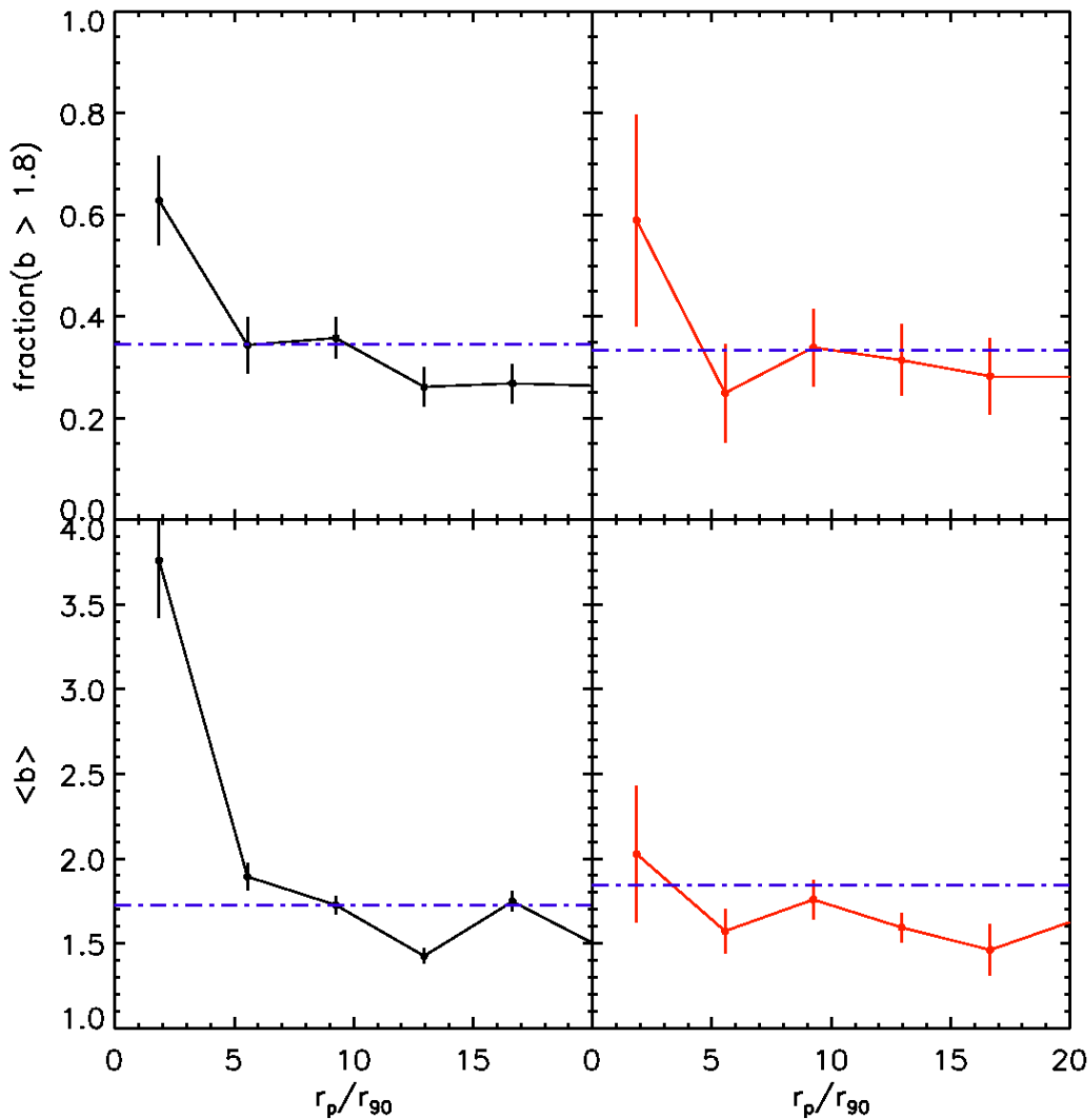}
\caption{Upper panel: $V/V_{max}$ weighted fraction of galaxies with significant
  stellar formation activity (birthrate parameter $b \ge 1.8$), as a
  function of the distance parameter to the nearest neighbor $r_p/r_{90}$. Lower
  panels: the $V/V_{max}$ weighted value of $<b>$ for galaxies, as a function
  of $r_p/r_{90}$. Left panels are for
the HSB case and right panels
  for
the LSB case. See text for details and conclusions from this
  Figure. \label{b}}
\end{figure}

\begin{figure}
\plotone{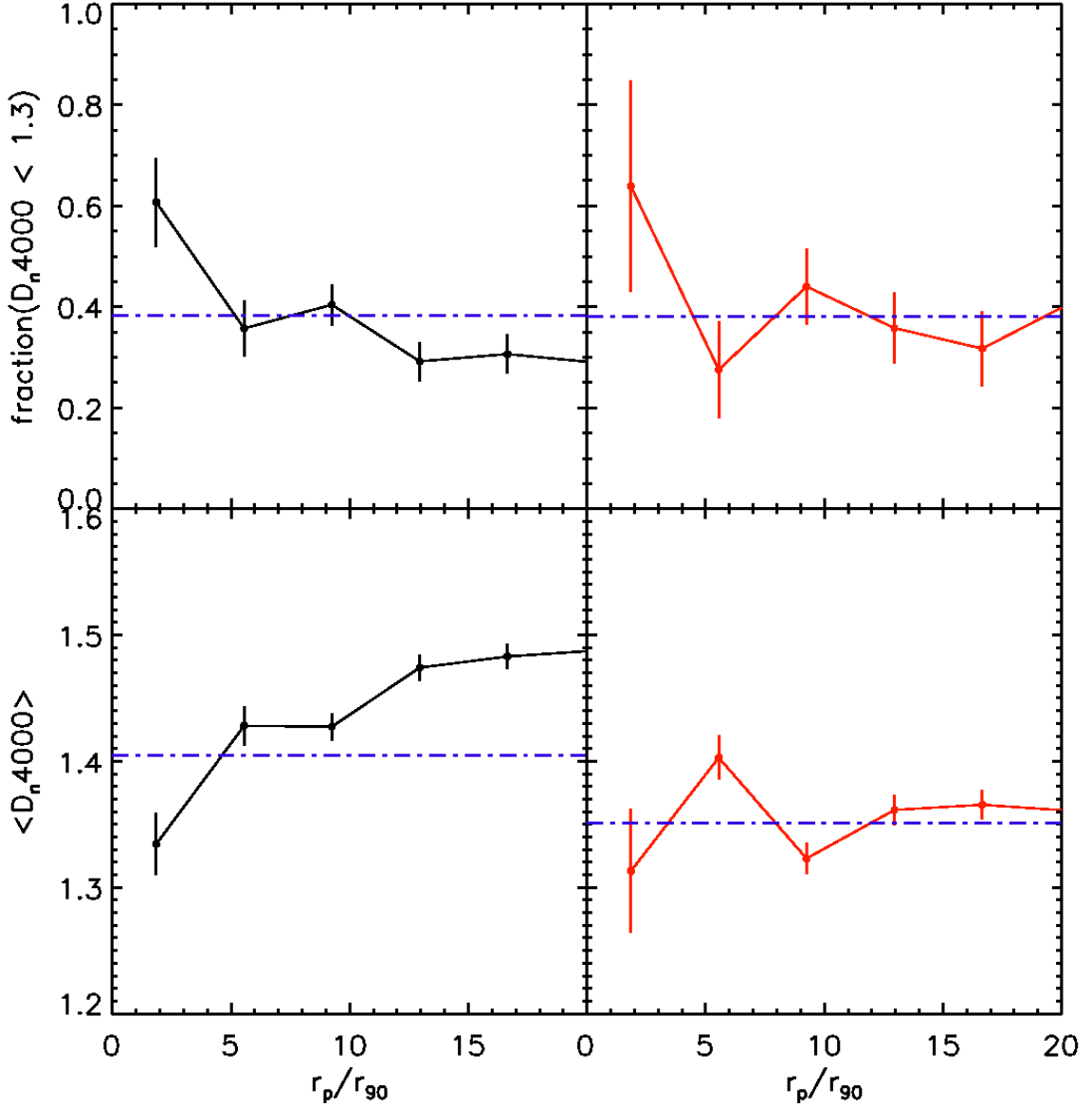}
\caption{Upper panel: $V/V_{max}$ weighted fraction of galaxies with
  very recently formed stars (D$_n$4000 index smaller than 1.3), as a
  function of the distance parameter to the nearest neighbor $r_p/r_{90}$. Lower
  panels: the $V/V_{max}$ weighted value of $<$D$_n4000>$ for
  galaxies, as a function 
  of $r_p/r_{90}$. The black lines denote the HSB case and the red
  lines the LSB case. See text for details and conclusions from this
  Figure. \label{b_dn4_rp}}
\end{figure}

Now that we ensure that the population of LSB and HSB pairs are
comparable, we can search for trends between the
star-forming parameters and the environment properties, such as the local
density. The 
upper panels of Figures \ref{b} and \ref{b_dn4_rp} show the distributions of
the birthrate parameter $b$ (Fig. \ref{b}) and the D$_n$4000 index
(Fig. \ref{b_dn4_rp}), for the sample of 
pairs and the control sample. Both distributions are defined above (respectively,
below for Fig. \ref{b_dn4_rp}) a threshold value from
which we can naturally define those systems with a high star-forming
activity (respectively, very recently formed stars for
Fig. \ref{b_dn4_rp}). In this case we choose pairs with $b > 1.8$ and
D$_n$4000 $< 1.3$. These systems, clearly with star-forming processes and/or
recently formed stars, are worth to be characterized in terms of the
projected distance to the closest neighbor $r_p$, normalized by the
Petrosian radius $r_{90}$. Results can be observed in lower panels of Figures \ref{b}
and \ref{b_dn4_rp}, where we plot the average birthrate parameter
$b$ and the D$_n$4000 index as a function of the $r_p/r_{90}$ ratio.
%In both Figures, the upper panel shows the fraction of galaxies, as a
%function of  $r_p/r_{90}$, having very rapid stellar formation
%processes ($b > 1.8$, for Figure \ref{b}), and having very recently
%formed stars (D$_n4000 < $1.3, for Figure \ref{b_dn4_rp}). 
The blue dot-dashed
lines represent the corresponding fractions and averages for the
control sample for both groups of galaxies. Note that in these two
figures, both the fractions and averages are weighted by
$V/V_{max}$. Figure \ref{b} shows an increase in the
stellar formation activity for HSBGs compared to the corresponding
control sample. This is in agreement with previous works
\citep[and references therein]{oneil2007}. Regarding
the LSBGs, both the control sample of LSBGs and the LSBGs 
with pairs, exhibit a significant fraction of galaxies with star
forming processes (around 30\%). This increases even more for pairs
with $r_p/r_{90} \le 4$, where the fraction of star-forming galaxies
doubles, to reach up to 60\%, which represents a large value when compared to
that for the control sample. Note that for the average birthrate parameter $b$,
$<b>$, we observe an increment in the first bin of  $r_p/r_{90}$,
twice as large as the increment in the same bin observed for the LSBG
pair sample. This behavior is mainly due to the fact that the $b$
distribution for the HSB galaxies dominates for $b > 4$, which {\em is not}
the case for the LSB galaxies (then the difference in the $<b>$ value
for HSBGs and LSBGs). 

On the other hand, the increase of the star formation signatures
described above is consistent with the fraction of young stars from
the D$_n$4000 index distribution (Figure 
\ref{b_dn4_rp}). Both samples of galaxies (LSBGs and HSBGs), exhibit an
increase in the fraction of galaxies having recent star 
formation episodes, when $r_p/r_{90}$ decreases, in comparison with the
corresponding control samples (the horizontal blue dot-dashed
lines). The smaller average values of $<$D$_n4000>$ for LSBGs compared
to HSBGs for $r_p/r_{90} > 10$, indicate that LSBGs in average have {\em
  younger} stellar 
populations than HSBGs at large separation scales. In fact, results
show that in average, the fraction of young stars in LSBGs is roughly
similar for all scales. Note that for both
LSBGs and HSBGs the average value for D$_n$4000 is smaller than the
corresponding value for the control sample.

\subsection{The AGN and LSB connection}
\label{agn}

There are several reasons for devoting special attention to the
fraction of LSBGs which host AGN activity. First, the evolution of an
AGN can alter the evolution of the entire galaxy \citep[and 
references therein]{silverman2008}. Second, it has been claimed
\citep{galaz2002} that the bulge size can be related to the
metallicity, where smaller bulges are metal poor, favoring a secular
evolution picture for the case of some barred spiral LSBGs and clearly
appealing to the mass metallicity relationship in non-barred spirals
\citep{lagos2009}. Concerning the fraction of 
LSBGs hosting AGNs, \citet{sprayberry1995b} found that {\em half} of his
sample of 10 giant LSBGs exhibit AGNs. The same fraction was also found
by \citet{schombert1998}. These rather high fractions were challenged
by \citet{impey2001}, who found that only 5\% of their LSB sample, selected
from the Automated Plate Measuring Machine Survey
\citep[APM]{maddox1990}, hosted AGNs. Finally, \citet{mei2009}, using 194 LSBGs 
from the spectroscopic catalog of \citet{impey1996}, along with
spectroscopic information from the SDSS DR 5, concluded that 10\%-20\% of
the LSBGs host an AGN, to be compared to the 40\%-50\% of HSBGs
presenting an AGN. 

%%%%%%%%%
The luminosity of the bulge has also an important role in the fraction
of spirals with an AGN. \citet{schombert1998} using a sample of
galaxies with high HI emission concluded that the occurrence of AGN was
highest in systems with bulges, regardless of morphological type or
mean surface brightness. However, the existence of a stellar
population (and the associated gravitational gradient) appears to be a
prerequisite for an active nuclear region. This evidence points to
other results indicating that the bulge and disk
evolution are decoupled and so whatever star formation processes
produced the bulges did not affect the disks \citep{das2009}.
%%%%%%

Motivated by the above, we devote attention to the HSBGs and
LSBGs hosting an AGN. The sample of galaxies hosting an AGN, for both
samples containing LSBGs and HSBGs, was obtained using the
\citet{kauffmann2003} catalog, using the well-known BPT diagnosis diagram
\citep{baldwin1981} from the intensity of the [OIII]$\lambda$5007,
H$\beta$, [NII]$\lambda$6583 and H$\alpha$ lines. Following the BPT
diagram, AGNs are defined as galaxies having

\begin{equation}
\log([OIII]/H\beta) \ge 0.61/(\log([NII]/H\alpha) - 0.05) + 1.3.
\end{equation} 

\begin{figure}
\plotone{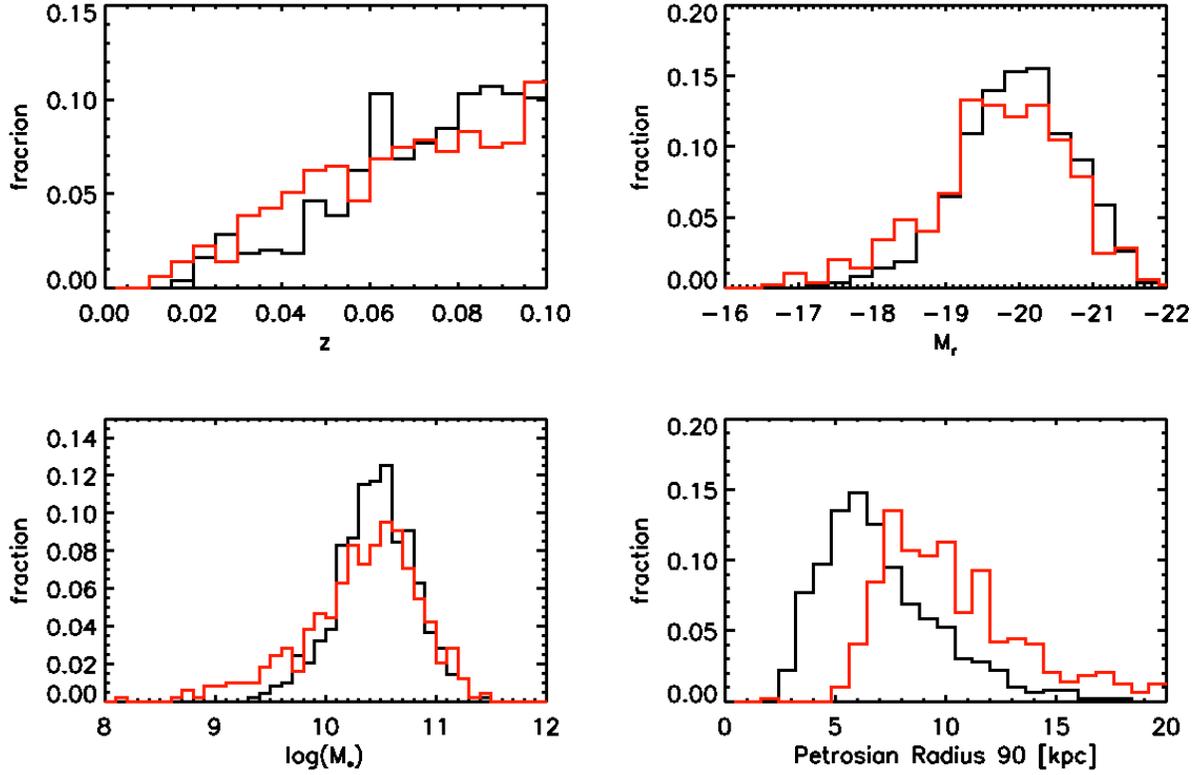}
\caption{Histograms for the LSBGs (red) and HSBGs (black) hosting an
  AGN. From the 
upper panel to the right to the bottom: redshift $z$,
  absolute magnitude in the $r$ band, log of the stellar mass, and the
  Petrosian radius $r_{90}$. Most of the 
distributions for LSBGs and HSBGs with
 AGN are similar, except the distribution for the sizes given by
  the Petrosian radius. See text for details. \label{histo_agn}}
\end{figure}

Using the above selection criterion, 495 out of 9,421 LSBGs present an
AGN. On the other
hand, 5,677 HSBGs out of a total of 30,000 also 
exhibit an AGN. In rough numbers then, 19\% of the HSBGs present an
AGN, compared to the 5\% of the LSBGs having an AGN.  
Figure \ref{histo_agn} shows the fraction of HSBGs and LSBGs in our
sample presenting
AGNs, as a function of the redshift, the absolute magnitude in the
$r$-band, the log of the stellar mass, and the Petrosian radius 
$r_{90}$ in kpc. Both in the HSB and LSB galaxies, we observe that the
redshift distributions {\em and} the absolute magnitude distributions
for both LSBGs and HSBGs are similar. This is expected since both AGNs
and LSBGs are detected in the brightest portion of absolute magnitude range. A
similar distribution is also observed in the stellar mass distribution
for both the LSBGs and HSBGs having AGNs (left bottom panel). Where
things are clearly different is in the size distribution of HSBGs and
LSBGs having an AGN (right bottom panel). Although both distributions
are similar in shape, LSBGs hosting an AGN are larger, in average, than the HSB
counterparts. Also, when comparing $r_{90}$ between Figure
\ref{histo_agn} and Figure \ref{gen_props} (or Figure
\ref{gen_props_corrected}), it is worth noting that both LSBGs and HSBGs hosting 
an AGN are {\em smaller} in average than the counterparts lacking an
AGN. We suggest that the presence of an AGN could be responsible for
this difference observed statistically.
%Could the presence of an AGN be held responsible for this
%difference observed statistically? 

\begin{figure}[!ht]
\plotone{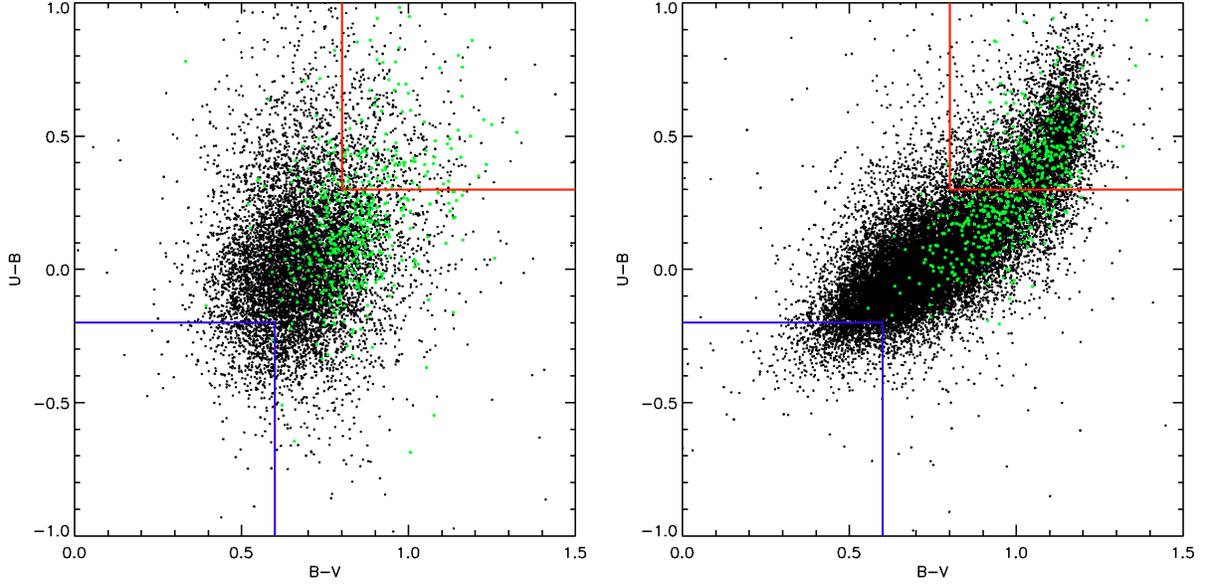}
\caption{$B-V$ vs. 
$U-B$ color-color diagram of galaxies classified as
  LSBGs (left) and as HSBGs (right) from the SDSS data. The red and blue lines indicate
  what we define as ``red'' and ``blue'' galaxies, respectively. Green
  points are AGN-LSBGs. See text for details. \label{color_ambas}}
\end{figure}

In Figure \ref{color_ambas} we compare the color properties of LSBGs
and HSBGs with and without AGN, via the $U-B$ vs. 
$B-V$ color-color diagram
(LSBGs in the left panel, and HSBGs in the right panel). Black points
represent the galaxies without AGN  
and the green points represent galaxies hosting an AGN. For both HSBGs
and LSBGs, galaxies having an AGN are in average redder than their
counterparts lacking an AGN. The average values for
LSBGs having an AGN is $U - B= 0.23\pm0.01$ and $B - V = 0.85\pm0.01$. This
locates the LSBGs-AGN galaxies in the space covered mainly by the red
population of LSBGs (delimited by the red line), 
 a region first noted by
\citet{oneil1997}, and reinforced by \citet{galaz2002} who show, using
near-IR photometry, that a large fraction of LSBGs with high gas
fractions host old and red stellar populations. 

\begin{figure}
\plottwo{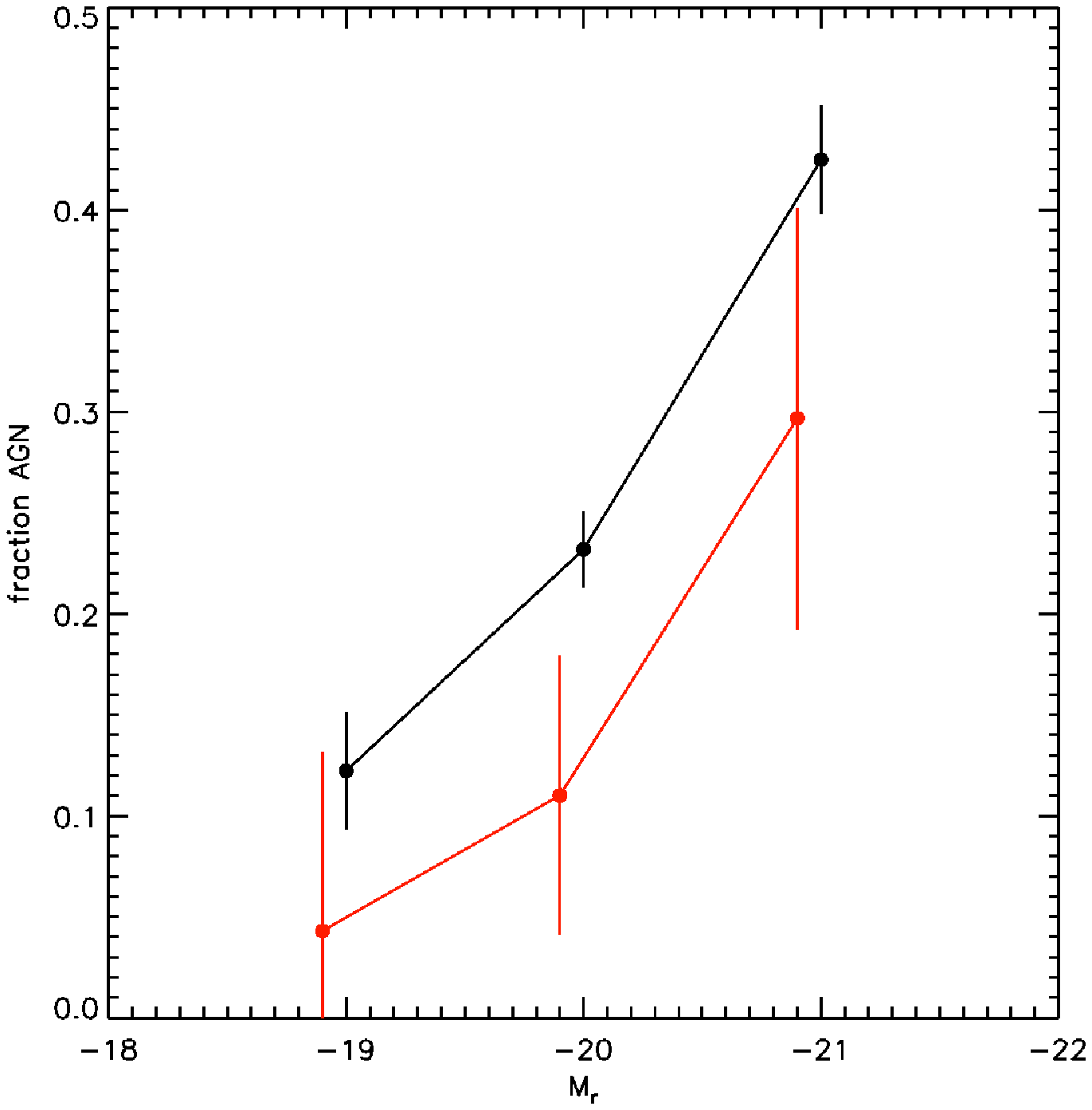}{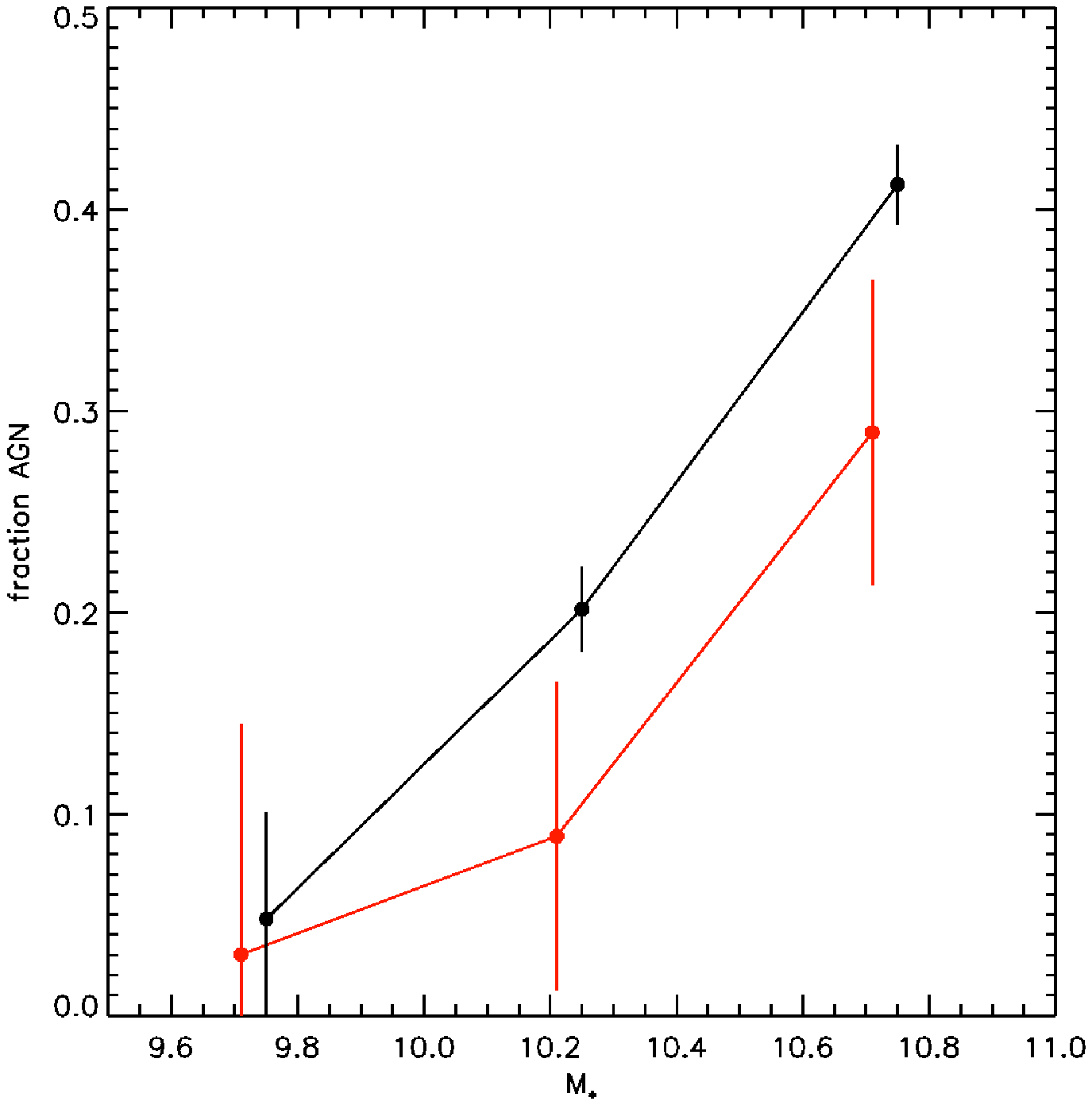}
\caption{Fraction of HSBGs (black) and LSBGs (red) hosting an AGN, as a
  function of the absolute magnitude in the $r$ band (left) and the
  stellar mass (right). For clarity, the x-axis of the LSBG sample has
  been slightly shifted to the left.  \label{frac_agn}} 
\end{figure}

The fraction of LSBGs and HSBGs with AGN, as a function of the absolute
magnitude and the log of the stellar mass, is shown in Figure
\ref{frac_agn}. In agreement with the previous authors already cited,
the fraction of LSBGs hosting an AGN is always lower than the
corresponding fraction of HSBGs also hosting an AGN, regardless of the
absolute magnitude and the stellar mass. It is not surprising that both
parameters evolve in the same way (larger fraction of AGNs for brighter
objects and larger masses), since both quantities are related via the
mass-luminosity relation. As this Figure shows, the AGN fraction
is a steep function of the brightness (and the mass) of the 
galaxy, as observed by many authors \citep[and references
therein]{hopkins2007}. As it has been claimed, this 
is just another realization that the mass of the black hole (and then
the power of the AGN), is related to the mass of the galaxy itself
\citep[and references therein]{marconi2003}. For the lowest covered masses
($\sim 10^{9.8}$ M$_\odot$), 
the fraction of LSBGs hosting an AGN is $\sim 5\%$, roughly the same
fraction observed for the HSBGs. However, for the larger masses covered
by this statistics, the fraction of LSBGs hosting an AGN reaches $\sim
30\%$, which is 10\% {\em lower} than the corresponding fraction of HSBGs hosting an
AGN. These 
fractions are quite similar when using the absolute
magnitude. Why, regardless the stellar mass of a galaxy, the fraction
of LSBGs hosting an AGN is lower than the fraction of HSBGs having AGNs? This
may be related to the star formation properties of LSBGs in
different environments described in \S\ref{interactingLSBGs}, a subject
we discuss in \S\ref{discussion}. 

\begin{figure}
\plotone{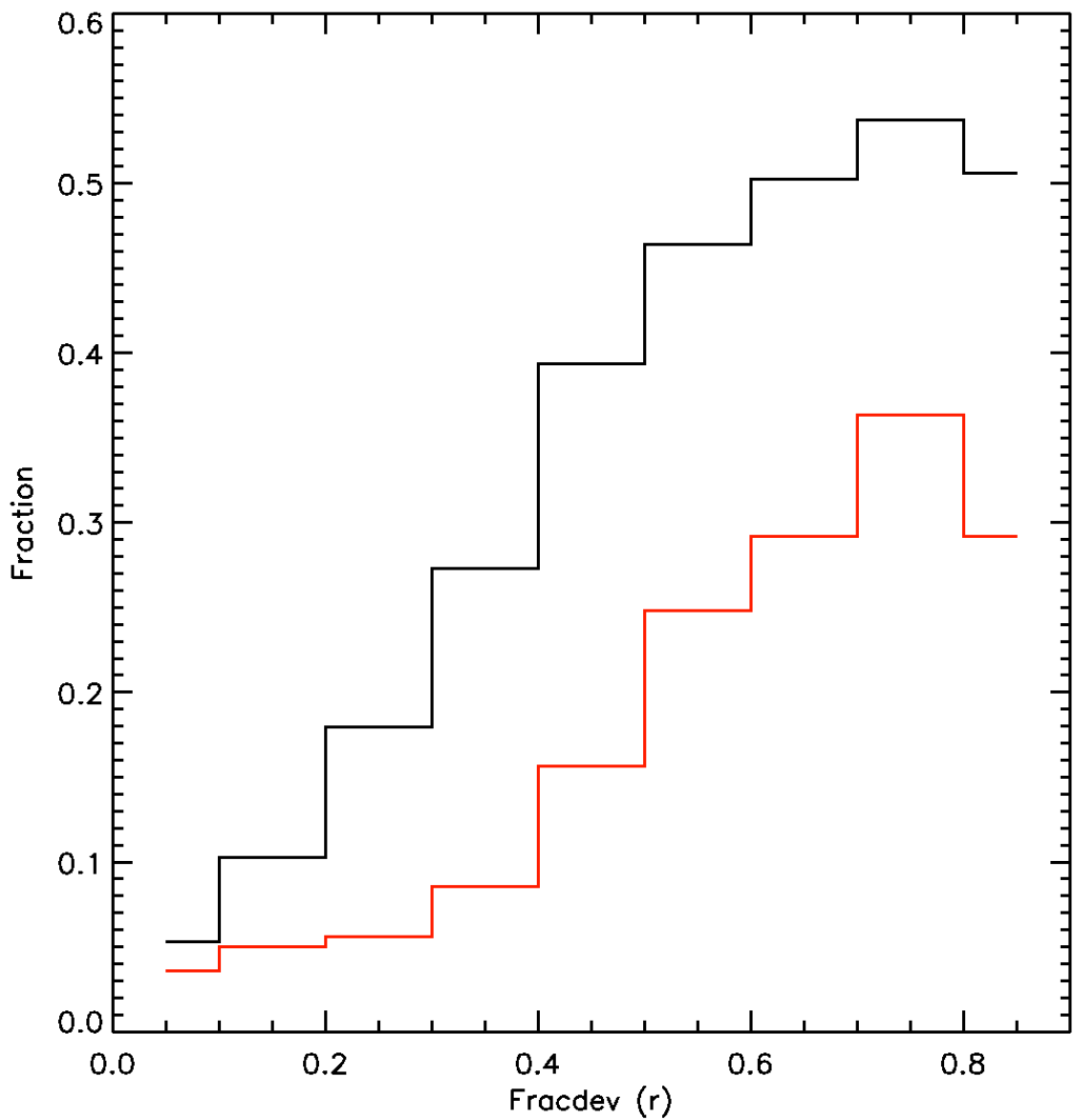}
\caption{The fraction of HSBGs (solid line) and LSBGs (dashed line)
  having AGN as a function of the FRACDEV parameter, which represents the
  luminosity contribution of the de Vaucouleurs profile (or bulge
  profile). For both the HSBGs and the LSBGs populations, the fraction
  of galaxies hosting an AGN increases when the
  luminosity of the bulge increases. The fraction for each bin of fracDev
  is obtained by dividing the number of galaxies with an AGN in the given
  population in that bin, by the total number of galaxies of the
  same population in the same fracDev bin.    \label{agn_bulge}} 
\end{figure}

\begin{figure}
\plotone{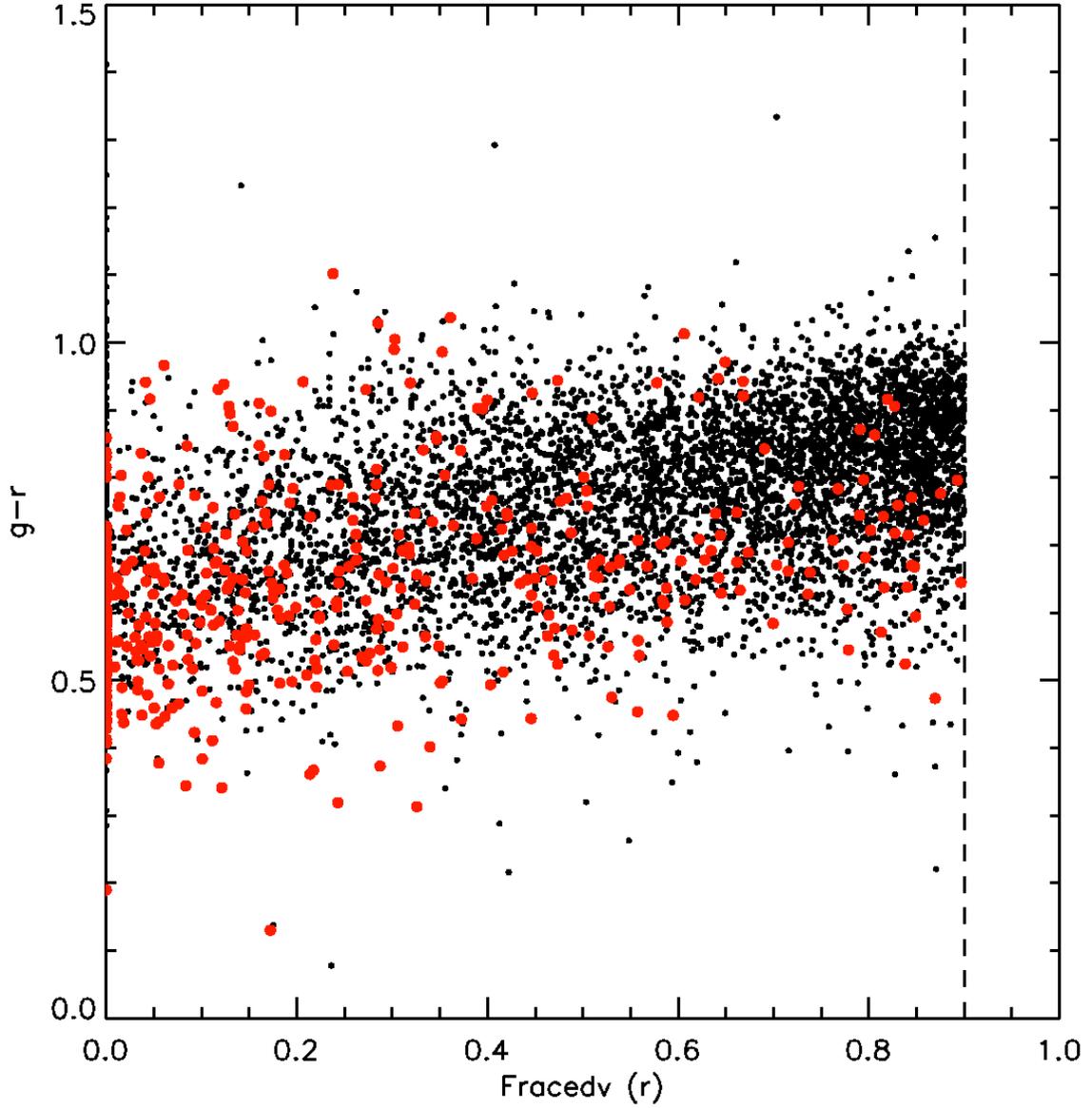}
\caption{The color $g - r$ of the HSBGs 
 (black) and LSBGs (red) hosting
  an AGN, as a function of the fracDev parameter representing the
  fraction of the total light given by the bulge. LSBGs tend to have
  smaller and less luminous bulges than HSBs. The vertical dashed
  line represents the limit in fracDev where the contamination by
  ellipticals begins to be significant.   \label{color-agn}} 
\end{figure}

%%%%%%%%%%
Regarding the bulge luminosity and the fraction of galaxies with AGN,
we obtain similar results for the two galaxy populations (HSBGs
and LSBGs). Figure \ref{agn_bulge} shows the fraction of LSBGs and
HSBGs with AGN as a function of the parameter used to define the luminosity
contribution of the bulge (i.e. the fracDev parameter), represented by a de
Vaucouleurs luminosity profile. Both distributions
are normalized by the total number of galaxies of the same population
found in each bin of fracDev. The black and red lines show the trend for the
HSBGs and the LSBGs, respectively. It is apparent that both fractions 
increase (with almost the same slope) as the bulge dominates the
galaxy luminosity, a result in 
agreement with other works discussed here \citep[and references
therein]{schombert1998}. We see also that the HSBG population has a
larger fraction of AGN compared to LSBG counterparts, in agreement
with the fact that the HSBGs host a larger fraction of AGN than the
LSBG counterparts (19\% for the HSBGs and 5\% for the LSBGs). In
order to check the morphology of the galaxies with AGN, we examined by
eye the HSBGs and LSBGs hosting an AGN. Both populations show
exclusively spiral galaxies with a visible bulge, the LSBGs-AGN
exhibiting bluer colors than the HSBG-AGN population and less
prominent bulges (see Figure \ref{color-agn}). However, the LSBG-AGN 
population has a larger fraction of small bulges. These galaxies have
spectra with less prominent AGN signatures compared to those with more
luminous bulges. All this evidence is in agreement with the trend
observed in other AGNs, where there is a close relationship between
the bulge luminosity and the power of the AGN, represented by the
emission line diagnosis.  
% One possibility is that the plot is just showing a selection
% effect for the LSB populations. For example, when LSBGs have a more
% luminous bulge, but still being a low surface brightness galaxy, less
% emission lines are present, resulting in a more frequent failure to
% measure redshift and then entering into the spectroscopic catalogue.
% However, colors of the LSBG-AGN population are bluer compared to the
% HSBG-AGN counterparts, with a value similar to the color difference
% between the LSBGs and HSBGs without AGN. This 
% translates in a higher fraction of blue galaxies with spiral
% morphologies, when compared with HSB-AGNs. This last affirmation is
% confirmed when we examine the LSBG-AGN and HSB-LSB images by
% eye. Interestingly, we observe that this turns into a higher fraction
% of galaxies exhibiting bars compared with the population of HSBGs with
% an AGN (about 40\% more). One can speculate 
% that when bars are present, then an AGN is more easily activated,
% which happens when the nucleus is not necessarily brighter/larger. In
% any case, the analysis of this interesting result worth more attention
% in a further study. 
%%%%%%%%%%%%%%

\section{Discussion}
\label{discussion}

As argued in the introduction, there is still a debate about the
evolution of LSBGs when located in dense environments. Are LSBGs
destroyed in dense environments, or they just change their condition
of LSBGs, 
 turning into HSBGs? Given that these galaxies dominate the
volume density of galaxies \citep{dalcanton1997}, then having clues to
solve this issue is of fundamental importance. There are basically two
different scenarios describing the evolution of an LSBG in dense
environments, especially when an LSBG is subject to the direct
influence of another galaxy. The first scenario establish that LSBG
disks are unstable under perturbations (given their low density), making LSBGs very
sensitive to the local density and hence making them to populate only
low density environments. A second scenario, also supported by models
\citep[hereafter MMB model]{mihos1997} and observations
\citep{pickering1997, das2010}, yields {\em stable disks 
  against interactions}, for 
sufficiently {\em low} stellar mass densities ($\Sigma_s$) and assuming
large amounts of dark matter, as it is the case for LSBGs. In this
picture, below a mass density threshold, perturbations generated by
the interactions would be unable to be amplified, preventing the galaxy
from being destroyed and/or producing significant star formation
episodes. The net result is a much more stable galaxy against close
interactions. Such a stability would be represented, for example, by
the lack of bars and the absence of flow of material which can form
stars or even feed a central massive black hole (activating an
AGN). Therefore, the deficit of LSBGs hosting AGNs, compared to 
those found in HSBGs, could fit at least with the picture the MMB
model presents.  

This last picture also fits well, but with a subtlety, with the
values of the stellar formation indexes $<b>$ and $<$D$_n$4000$>$ as a
function of the 
interacting distance $r_p/r_{90}$, as found for both LSBGs and
HSBGs (Figures \ref{b} and \ref{b_dn4_rp}). When comparing the fraction of LSBGs 
and HSBGs with the {\em highest} values of $b$ ($b > 1.8$), we realize
that these fractions are the same at all
scales, for the whole range of $r_p/r_{90}$; i.e., there is no
distinction in the star forming properties between LSBGs and HSBGs,
where the latter are actively forming stars. However, when looking at the complete
range of $b$, we discover that at smaller scales (i.e. close
interactions, $r_p/r_{90} < 5$), the fraction of HSBGs having stellar
formation episodes {\em double} that of LSBGs. These results could
match the two scenarios described 
above. In one case, the lack of LSBGs with high values of $b$
(significant stellar formation episodes) points to the possibility that
such galaxies could no longer be LSBGs, having evolved to a higher surface
brightness regime. On the other hand, following the MMB model, the
difference in the interaction signature could be explained by the disk
stability, preventing LSBGs from changing their star formation rate,
as it is observed in HSBGs. 

Thus, our results support the picture where LSBGs are more isolated
than HSBGs at large scales, a result in agreement with those
by \citet{rosenbaum2009}. However, more than the local
density value, are the interactions with close neighbors which actually make
LSBGs evolve. Interestingly, when this occurs, LSBGs seem to be more
stable than HSBGs, preventing LSBGs to change significantly their
stellar formation signatures as usually expected with respect to
HSBGs with close companions. This last behavior, together with the
lower fraction of LSBGs with 
AGNs compared to the fraction of HSBGs also hosting AGNs (for
the full mass range), could be explained by the difficulty shown by
interacting LSBGs to form bars, lacking the natural
bridge which transports material to the center of the galaxy,
preventing the central black hole from being fueled, and then
remaining the galaxy as a non-AGN.

\section{Conclusions}
\label{conclusion}

Our conclusions can be summarized as follows. 
\begin{itemize}
\item For a sample of 9,421 LSB galaxies selected from the SDSS DR 4
  (1.66\% of a total of 567,486 galaxies),
  in the redshift range $0.01 < z < 0.10$, the cumulative
  distributions of the distance to the nearest and the fifth closest
  neighbors, indicate that LSBGs are more isolated than high surface
  brightness galaxies (HSBGs). Also, LSBGs specially avoid high density
  environments, i.e. are almost completely {\em unpopular}, following
  the \citet{bothun1993} terminology. When describing the degree of isolation
  in terms of the surface density of galaxies, our results yield a deficit
  of neighbors for LSBGs at small scales ($<$1 Mpc). For larger scales
  ($\ge 2.5$ Mpc), we obtain consistent fractions of LSBGs and
  HSBGs. This is in fairly good agreement with results by 
  \citet{rosenbaum2009}. However, we do not detect a higher degree of
  isolation of LSBGs in the scale range 2-5 Mpc as these authors
  do. This difference can be explained by the different galaxy
  selection criteria, as most of our galaxies are spirals, losing some
  spirals with extremely bright bulges {\em 
    and} faint disks, as well as faint irregulars ($M_r > -16.0$ mag),
  and Rosenbaum's low redshift ones are mostly dwarfs.  
%  as an straightforward explanation is not evident. 
\item The fraction of LSBGs with high stellar formation signatures 
  and/or a high population of recently formed stars, increases when
  galaxies are close to a neighbor ($r_p/r_{90} \le 4$), and is clearly
  larger when compared to a sample of isolated LSBGs. This behavior is
  similar in HSBGs also suffering interactions. However, when
  comparing the average value of the stellar formation signature
  (e.g. the birthrate parameter $b$), interacting HSBGs form
  {\em twice} the stars as the interacting LSBGs. This last difference
  could be explained by models of \citet[MMP models]{mihos1997}, which
  show the difficulty of low density stellar disks (i.e. low surface brightness
  disks) surrounded by significant dark matter haloes, to amplify and
  propagate perturbations due to close companions.  
\item The fraction of LSBGs hosting an AGN is lower than the fraction
  of HSBGs with AGNs. 5\% of the LSBGs
  host an AGN, compared to the 
% 32\% of the HSB systems. These two
  19\% of the HSB systems. These two
  fractions are in agreement to those found by 
%\citet{mei2009}. The
  \citet{impey2001}. This is systematic for the whole range of
  masses covered in this work (i.e. for the whole range of absolute
  magnitudes of the volume-limited sample, $-22.0 < M_r <
  -19.8$). This could be explained by the difficulty of LSBGs to
  react under close companions (MMP models), which would produce a deficit of bars
  and other structures in LSBGs capable to transport material to
  activate the massive central black hole, which when active, would generate
  an AGN. Another interesting result is obtained when the frequency of
  an AGN is compared between the LSBG and HSBG populations, in terms of
  the bulge optical luminosity. We obtain that for both the HSBG and
  the LSBG populations, the fraction of AGNs increases with the bulge
  luminosity with almost the same slope, in agreement with other authors. 
  % However, for the LSBGs, the number of systems
  % exhibiting an AGN, decreases when the bulge luminosity
  % increases. Observing these two populations by eye, we discover that
  % as the nucleus of LSBGs with AGNs gets fainter (and smaller), the
  % fraction of spirals with bars also increases, a feature not observed
  % in the HSBG population. Although there could be some selection
  % biases, we suggest that this result should be studied in more
  % detail, in a separate work.  
%18\% of the LSBGs 
\end{itemize}

As a general conclusion, this study suggests that, rather than being a
condition for their formation, isolation of LSBGs is
more connected to  their survival and evolution.

\acknowledgments

This research is supported by FONDECYT grant
1085267, by FONDAP Center for Astrophysics 15010003, and BASAL
CATA. GG thanks the Department of Astronomy, University of 
Washington, Seattle, and the Department of Astronomy of the University of
Maryland, at College Park, for their welcome during the writing of
this paper. In particular, GG acknowledges Julianne Dalcanton for the fruitful
discussions on the subject covered by this work. The authors are
grateful to the anonymous referee who helped to improve the
form and content of this paper. 
 
Funding for the SDSS and SDSS-II has been provided by the 
Alfred P. Sloan Foundation, the Participating Institutions, the
National Science Foundation, the U.S. Department of Energy, the
National Aeronautics and Space Administration, the Japanese
Monbukagakusho, the Max Planck Society, and the Higher Education
Funding Council for England. The SDSS Web Site is
http://www.sdss.org/. The SDSS is managed by the Astrophysical
Research Consortium for 
the Participating Institutions. The Participating Institutions are the
American Museum of Natural History, Astrophysical Institute Potsdam,
University of Basel, University of Cambridge, Case Western Reserve
University, University of Chicago, Drexel University, Fermilab, the
Institute for Advanced Study, the Japan Participation Group, Johns
Hopkins University, the Joint Institute for Nuclear Astrophysics, the
Kavli Institute for Particle Astrophysics and Cosmology, the Korean
Scientist Group, the Chinese Academy of Sciences (LAMOST), Los Alamos
National Laboratory, the Max-Planck-Institute for Astronomy (MPIA),
the Max-Planck-Institute for Astrophysics (MPA), New Mexico State
University, Ohio State University, University of Pittsburgh,
University of Portsmouth, Princeton University, the United States
Naval Observatory, and the University of Washington.

% Figures...

%% the file, uncomment the \plotone, \plottwo, and
%% If you need a layout that cannot be achieved with \plotone or
%% \plottwo, you can invoke the graphicx package directly with the
%% \includegraphics command or use \plotfiddle. For more information,
%% please see the tutorial on "Using Electronic Art with AASTeX" in the
%% documentation section at the AASTeX Web site,
%% http://www.journals.uchicago.edu/AAS/AASTeX.
%%
%% The examples below also include sample markup for submission of
%% supplemental electronic materials. As always, be sure to check
%% the instructions to authors for the journal you are submitting to
%% for specific submissions guidelines as they vary from

%% This example uses \plotone to include an EPS file scaled to
%% 80% of its natural size with \epsscale. Its caption
%% has been written to indicate that additional figure parts will be
%% available in the electronic journal.

\clearpage

%% Here we use \plottwo to present two versions of the same figure,
%% one in black and white for print the other in RGB color
%% for online presentation. Note that the caption indicates
%% that a color version of the figure will be available online.
%%

%\begin{figure}
%\plotone{low_z.eps}
%\caption{The fraction of HSBGs (solid line) and LSBGs (dashed line) as
%  a function of the Petrosian radius $r_{90}$ for the galaxies located
%  at $z \le 0.01$. The total numbers are LSBGs 741, and HSBGs
%  1099. \label{low_z}}
%\end{figure}


%%% Tables....

%% If you are not including electonic art with your submission, you may
%% mark up your captions using the \figcaption command. See the
%% No more than seven \figcaption commands are allowed per page,
%% so if you have more than seven captions, insert a \clearpage

%% Tables should be submitted one per page, so put a \clearpage before

%% Two options are available to the author for producing tables:  the
%% deluxetable environment provided by the AASTeX package or the LaTeX
%% table environment.  Use of deluxetable is preferred.
%%

%% Three table samples follow, two marked up in the deluxetable environment,
%% one marked up as a LaTeX table.

%% In this first example, note that the \tabletypesize{}
%% command has been used to reduce the font size of the table.
%% We also use the \rotate command to rotate the table to
%% landscape orientation since it is very wide even at the
%% Note also that the \label command needs to be placed

%% This table also includes a table comment indicating that the full
%% version will be available in machine-readable format in the electronic

\begin{deluxetable}{lrrrr}
\tablecolumns{5}
\tablewidth{0pc}
\tablecaption{Averaged quantities for the $V/V_{max}$ and volume-limited
  samples of LSBGs and HSBGs. \label{stat_props}}
\tablehead{
\colhead{} & \multicolumn{2}{c}{$V/V_{max}$ catalog} & \multicolumn{2}{c}{Volume-limited sample} \\
\cline{2-3} \cline{4-5} \\
\colhead{} & \colhead{HSB} & \colhead{LSB} & \colhead{HSB} & \colhead{LSB}}
\startdata
$z$\tablenotemark{1}   & $0.063\pm0.001$ & $0.051\pm0.002$ & $0.077\pm0.001$ & $0.079\pm0.001$ \\
M$_r$\tablenotemark{2} & $-19.29\pm0.01$ & $-18.48\pm0.01$ & $-20.29\pm0.01$ & $-20.22\pm0.01$ \\
M$_*$\tablenotemark{3,4} & $9.96\pm0.08$   & $9.46\pm0.01$   & $10.44\pm0.01$  & $10.43\pm0.03$  \\
SFR\tablenotemark{5,4}   & $0.15\pm0.01$   & $-0.46\pm0.02$  & $0.52\pm0.02$   & $-0.39\pm0.03$  \\
$b$\tablenotemark{6,4}     & $0.034\pm0.01$  & $-0.08\pm0.02$  & $-0.08\pm0.01$  & $-0.19\pm0.02$  \\
r$_{90}$\tablenotemark{7} &$7.48\pm0.02$   & $8.99\pm0.03$   & $10.19\pm0.03$ & $14.24\pm0.09$ \\
Numb. Galaxies\tablenotemark{8}         & 30,000         & 9,421           &  7,526         &  1,110 \\
Numb. AGN\tablenotemark{9}              & 5,677          & 495             & \nodata        & \nodata \\
\enddata
\tablenotetext{1}{Redshift.}
\tablenotetext{2}{Absolute magnitude in the $r$ band.}
\tablenotetext{3}{Stellar mass, in solar masses (M$_\odot$).}
\tablenotetext{4}{ Logarithmic value (base 10).} %NELSON: no salia la L en el ps, puse un espacio
\tablenotetext{5}{Star formation rate (M$_\odot$/yr).}
\tablenotetext{6}{Birthrate parameter, expressed in star formation rate per unit of solar mass.}
\tablenotetext{7}{Petrosian radius in kpc in the $r$-band, which encompasses 90\% of 
the galaxy light.}
\tablenotetext{8}{Number of galaxies in each sample.}
\tablenotetext{9}{Number of galaxies of the corresponding sample,
  presenting an AGN. See text for details.}
\end{deluxetable}

\begin{deluxetable}{lrrrr}
\tablecolumns{3}
\tablewidth{0pc}
\tablecaption{Some averaged values for {\em isolated} and {\em popular} LSBGs and HSBGs in the sample.\label{props_sfr}}
\tablehead{
\colhead{} & \multicolumn{2}{c}{Isolated} &
\multicolumn{2}{c}{Popular} \\
\cline{2-3} \cline{4-5} \\
\colhead{} & \colhead{HSBGs} & \colhead{LSBGs} & \colhead{HSBGs} & \colhead{LSBGs} \\
} 
\startdata
M$_r$\tablenotemark{1}     & $-20.35\pm0.01$ & $-20.32\pm0.07$ & $-20.34\pm0.01$ & $-20.21\pm0.06$ \\
$g - r$\tablenotemark{2}   & $0.68\pm0.02$ & $0.63\pm0.02$   & $0.52\pm0.05$ & $0.59\pm0.01$ \\
$b$\tablenotemark{3,4}     & $1.18\pm0.04$ & $1.1\pm0.1$     & $1.29\pm0.04$ & $0.9\pm0.1$ \\
D$_n$4000\tablenotemark{5} & $1.55\pm0.01$ & $1.57\pm0.03$   & $1.53\pm0.01$ & $1.54\pm0.03$ \\
\enddata
\tablenotetext{1}{Absolute magnitude in the $r$-band.}
\tablenotetext{2}{$g - r$ rest-frame color, from the SDSS DR 4.}
\tablenotetext{3}{Birthrate parameter, expressed in star formation
  rate per unit of solar mass.} 
\tablenotetext{4}{ Logarithmic value (base 10).} %NELSON: no salia la
                                %L en el ps, puse un espacio 
\tablenotetext{5}{D$_n$4000 index.}
\end{deluxetable}

\end{document}